\documentstyle[12pt,psfig]{article}
\begin{document}

\input epsf
\newcommand{\sqtt}{\mbox{$\sqrt{3} \times \sqrt{3}$}}
\newcommand{\qz}{\mbox{$q=0$}}
\newcommand{\qzp}{\mbox{$q=0(++)$}}
\newcommand{\qzm}{\mbox{$q=0(--)$}}
\newcommand{\tkt}{\mbox{$T_{KT}$}}
\newcommand{\tktot}{\mbox{$T_{KT,1/3}$}}
\newcommand{\msqttsq}{\mbox{$|m_{\sqrt{3} \times \sqrt{3}}|^2$}}
\newcommand{\mqzsq}{\mbox{$|m_{q=0}|^2$}}
\newcommand{\xiasq}{\mbox{$(\frac{\xi_0}{a})^2$}}
\newcommand{\tdxiasq}{\mbox{$(\frac{\tilde{\xi}_0}{a})^2$}}
\newcommand{\tktsqtt}{\mbox{$T_{KT,\sqrt{3} \times \sqrt{3}}$}}
\newcommand{\tktqz}{\mbox{$T_{KT,q=0}$}}

\title{{\bf Superconducting Phase with Fractional Vortices
in the Frustrated Kagome Wire Network at $f=1/2$}}
\author{Kyungwha Park and  David A. Huse \\
{\small \it Department of Physics, Princeton University, Princeton, NJ 08544} }
\date{{\small \today}}
\maketitle

\begin{abstract}
In classical $XY$ kagome antiferromagnets, there can be a novel 
low temperature phase where $\psi^3=e^{i3\theta}$ has 
quasi-long-range order but $\psi$ is disordered, as well as 
more conventional antiferromagnetic
phases where $\psi$ is ordered in various possible 
patterns ($\theta$ is the angle of orientation of the spin). 
To investigate when these phases exist in a physical system,
we study superconducting kagome wire networks in a transverse
magnetic field when the magnetic flux through an elementary triangle
is a half of a flux quantum.  Within Ginzburg-Landau theory, we
calculate the helicity moduli of each phase to estimate the
Kosterlitz-Thouless (KT) transition temperatures. Then at the 
KT temperatures, we estimate the barriers to move vortices and
effects that lift the large degeneracy in the possible $\psi$
patterns. The effects we have considered are inductive couplings,
non-zero wire width, and the order-by-disorder effect due to
thermal fluctuations.  The first two effects prefer $q=0$
patterns while the last one selects a $\sqrt{3}\times\sqrt{3}$
pattern of supercurrents.  Using the parameters of 
recent experiments, we conclude
that at the KT temperature, the non-zero wire width effect dominates,
which stabilizes a conventional superconducting phase with a 
$q=0$ current pattern.  However, by adjusting the
experimental parameters, for example by bending the wires a
little, it appears that the novel $\psi^3$ 
superconducting phase can instead be 
stabilized.  The barriers to 
vortex motion are low enough that the system can equilibrate
into this phase.
\end{abstract}

\section{Introduction}

\begin{figure}
\begin{center}
\leavevmode
\epsfysize=5cm
\epsfbox{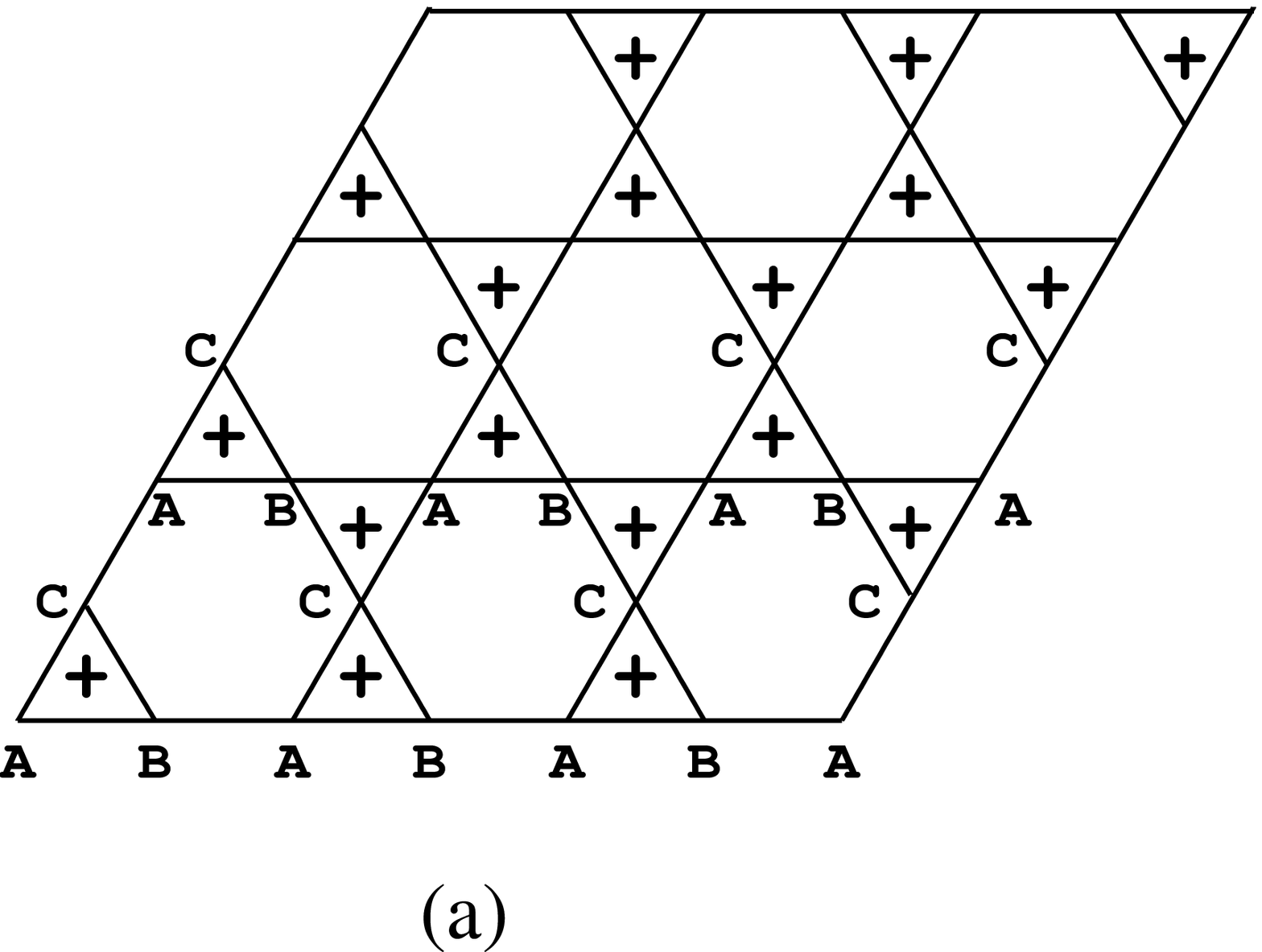}
\epsfysize=5cm
\epsfbox{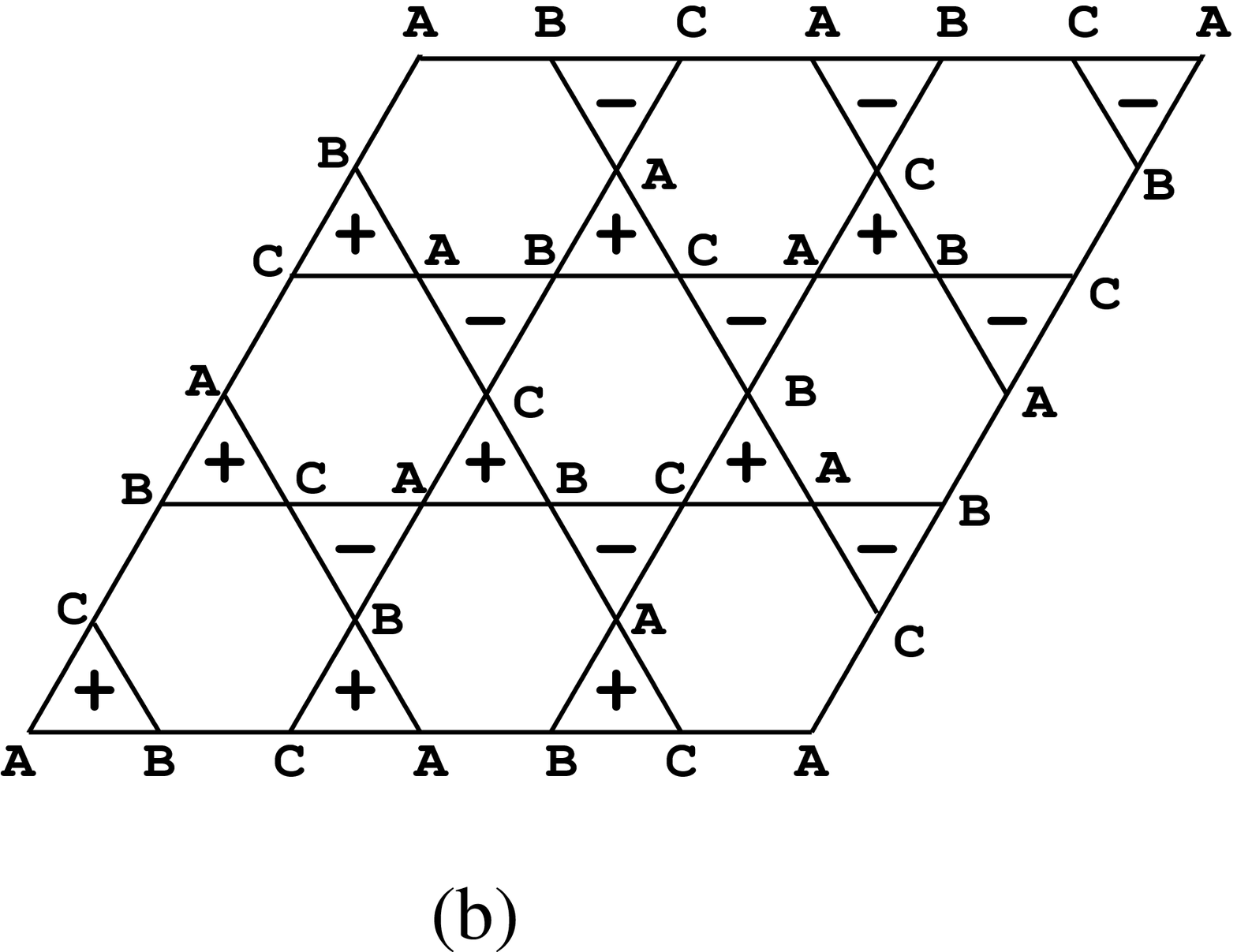}
\vspace{1cm}
\epsfysize=8cm
\epsfbox{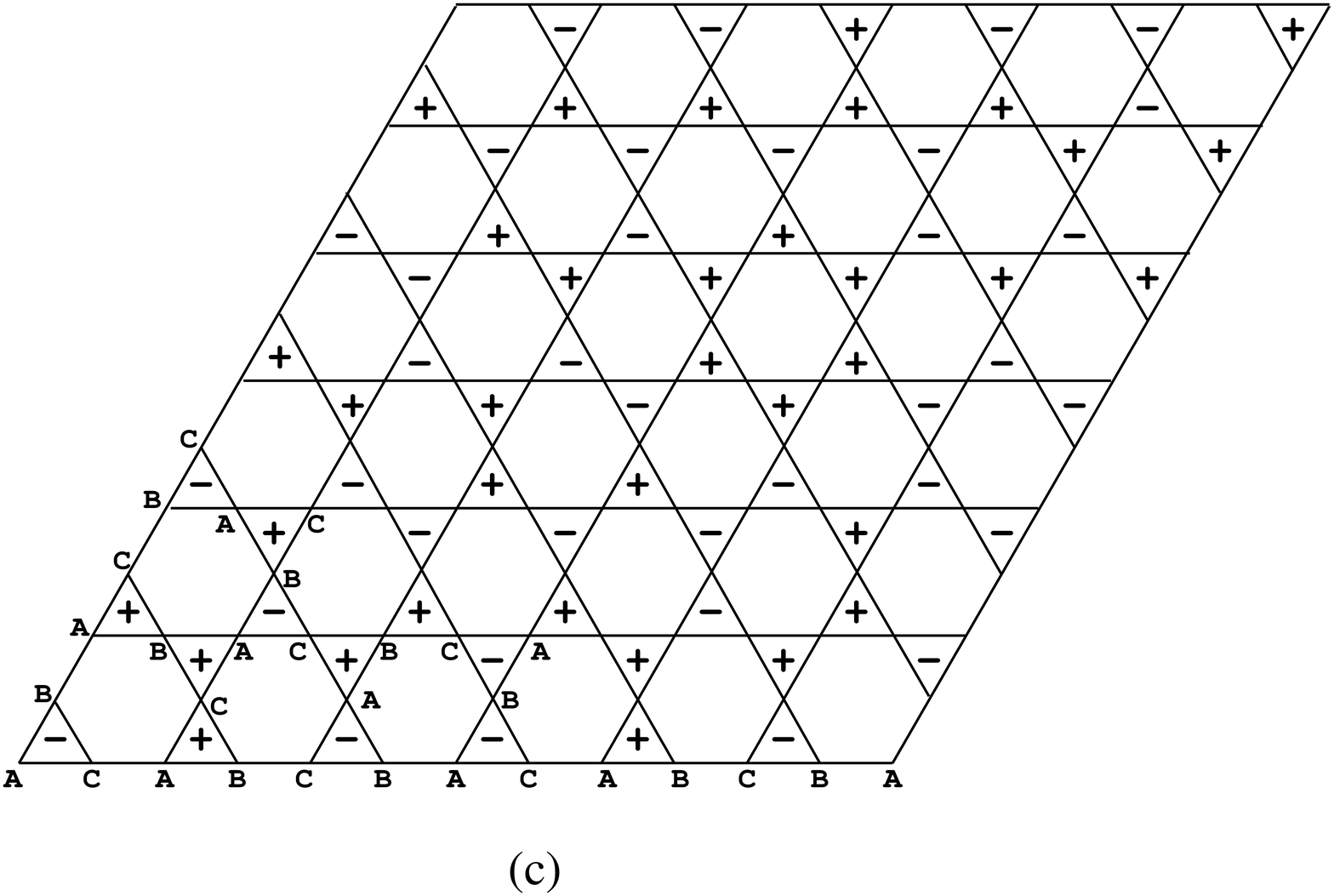}
\vspace{1cm}
\caption{Some ground state patterns of the classical
$XY$ kagome antiferromagnet.
Periodic boundary conditions are applied to these finite lattices.
$A$, $B$, and $C$ are the three possible spin orientations
at each site, $\theta = 0,\pm 2\pi/3$.  
When, on moving counterclockwise around a triangle,
the spins rotate 
counterclockwise
(clockwise) in spin space, a $+$ ($-$) sign is assigned to the triangle.
(a) One of the two $q=0$ patterns.
(b) One of the two $\sqrt{3}\times\sqrt{3}$ patterns.
(c) One of the random patterns.}
\label{antiferr-gs-patterns}
\end{center}
\end{figure}

Consider the classical $XY$ 
antiferromagnet on the kagome
lattice, with Hamiltonian 
\begin{eqnarray}
{\cal H} = -J\sum_{<ij>}cos(\theta_i-\theta_j)~, \label{eq:antiferr}
\end{eqnarray}
where 
$J < 0$, the sum runs over all
nearest neighbor pairs of sites, 
and $\theta_i$ is the orientation of the spin at site $i$.
Because of the antiferromagnetic interactions, 
each elementary triangular plaquette is frustrated.
The ground states of this system all have the angle 
$|\theta_i-\theta_j|=2\pi/3$ between all pairs of
nearest-neighbor spins.  This results in just three different
spin orientations, which we may call
$\theta=0,\pm 2\pi/3$, being present in the entire lattice.
All three spin orientations are present on 
every elementary triangle of the lattice.
Some examples of ground states are shown in Fig. \ref{antiferr-gs-patterns}.
There are two types of simply periodic ground states that
we will be particularly looking at.  One type is ordered with
zero crystal momentum, meaning the spin pattern has the same
translational symmetries as the lattice.  This type of ground state
is called ``$q=0$'' \cite{harris} and is shown 
in Fig. \ref{antiferr-gs-patterns} (a). 
The other simple ground state is called
``$\sqrt{3}\times \sqrt{3}$'' (Fig. \ref{antiferr-gs-patterns} (b)).
This pattern has a unit cell which consists of three
hexagons and which has a linear dimension $\sqrt{3}$ times larger than 
the grid's unit cell, thus its name \cite{harris}.
Unlike these simple patterns, almost all ground
states are instead irregular and
nonperiodic.  We call them ``random'' patterns.
One of these random patterns is shown in Fig. \ref{antiferr-gs-patterns} (c).
Because of the massive discrete degeneracy of the ground state patterns of 
$\psi=e^{i\theta}$, this system has an extensive ground-state 
entropy of roughly $0.126 k_B$ per site \cite{baxter}, in addition
to the usual continuous rotational degeneracy of the $XY$ spins.
This frustration-induced extensive degeneracy leads to a novel low-temperature 
ordered phase \cite{huse}\cite{rzchowski}.  In the novel low temperature phase,
$\psi^3=e^{i3\theta}$ has quasi-long range order but the
usual order parameter $\psi$ has only short-range
correlations.  Since the model is a two-dimensional $XY$ model, 
the low temperature phase appears only below a
Kosterlitz-Thouless (KT) transition temperature. \cite{kosterlitz}~
The vortices
that unbind at this KT transition are vortices of $\psi^3$,
and thus fractional 1/3-vortices in $\psi$.

Physical systems that realize frustrated $XY$ models 
are superconducting Josephson junctions or wire networks
on the kagome lattice in a transverse magnetic field.
In these systems the different spin patterns of the magnetic model
are realized as patterns of supercurrents.
In the kagome-lattice antiferromagnet the elementary 
triangular plaquettes are maximally frustrated, which 
occurs for the kagome superconducting Josephson junctions 
or wire networks when the magnetic flux through each 
elementary triangular plaquette 
is one-half of a flux quantum (or any odd multiple thereof). 
For a uniform field that maximally frustrates the triangular plaquettes,
the resulting flux through the hexagonal plaquettes of the kagome lattice
is an integer number of flux quanta, which means these plaquettes
are not frustrated, just as in the $XY$ antiferromagnet.  The novel
phase of interest does not have conventional superconductivity, with
quasi-long-range order (QLRO) in the phase $\theta$ of the Cooper pairs,
but does have QLRO in the phase $3\theta$ of ``Cooper sextuples''.
It is in this sense a rather exotic form of superconductivity.

Recently, an exciting experimental result \cite{yi}\cite{higgins} has emerged 
from aluminum kagome wire networks when the applied magnetic flux per triangular 
plaquette is one half of a flux quantum.  The result is that an ordered 
phase is appearing at this magnetic field, in contrast to 
the mean field prediction \cite{yn}.  
Motivated by these experimental data, in this paper we study 
this novel $\psi^3$ phase and other more conventional ordered phases
in kagome superconducting wire networks. 
However, the superconducting wire network does not {\it precisely} 
realize the model (\ref{eq:antiferr}), because (i) the material and 
connections do not consist of perfect mathematical points and straight
lines, (ii) there are also long-range inductive interactions between 
the supercurrents, and (iii) both
the {\it magnitude} and the phase of the superconducting order parameter
can vary near the temperatures of interest.  
All these (and possibly other) effects lift the
precise degeneracy that underlies the novel $\psi^3$ phase,
meaning the question of whether or not it can actually be realized
experimentally requires some more investigation.  This is the topic
of this paper.  We conclude that for the parameters of the
recent experiments (as summarized
in Appendix A) the novel $\psi^3$ ordered phase does not occur;
instead the system orders into a $q=0$ supercurrent pattern
with conventional superconducting order.  But the novel $\psi^3$
phase can be stabilized if one adjusts some experimental parameters
such as by bending the wires a little or making the wires narrower.

\section{Model and its Possible Superconducting Phases}

We model the superconducting wire array with the usual Ginzburg-Landau (GL)
free energy functional:
\begin{eqnarray}
F & = & \int d^3 x [ \alpha|\psi|^2 + \frac{\beta|\psi|^4}{2}
+ \frac{1}{2m^{\ast}} |(\frac{\hbar}{i}\vec\nabla - \frac{e^{\ast}}{c}\vec A)
 \psi|^2 ]~, \label{eq:orig-GL-F}
\end{eqnarray}
where the integral runs over the volume occupied by the wires.
The wires form a kagome grid, with distance $a$ 
between junctions.  The wire pattern is invariant under
translations by $2a$ parallel to any wire, and also has all the
point group symmetries of the ideal kagome lattice.  At first we
will assume the wires are straight and quite thin, with a rectangular
cross-section of in-plane width $w$ and thickness $d$.
[Later we will explicitly consider effects due to the nonzero width
of the wires and due to possibly bending the wires away from
straight.]  In this thin-wire limit, any variation of the order
parameter or vector potential across the wires is neglected, so the
model treats the wires as ideal lines and the junctions as points.
At this point we also neglect the energy of the magnetic fields
produced by the supercurrents.  These inductive effects are
estimated below and are negligible near the phase transition for
the experimental parameters we will consider here.   In zero magnetic 
field, the mean field phase transition temperature, $T_c^m$,
is the temperature where $\alpha = 0$.  We will be working
below but generally quite near this temperature so will assume, 
as usual, that $\alpha$ varies linearly with the temperature $T$.
We will neglect any temperature dependence of the other parameters
in the free energy.

For convenience, we will use $a$, the distance between wire
junctions, as our unit of length.  In zero field the free energy is
minimized when the magnitude of the order parameter $|\psi|$ is
$\sqrt{\frac{|\alpha|}{\beta}}$ and we will use this as the unit
for the order parameter.  We use $\frac{\Phi_0}{2\pi a}$ as the
unit of vector potential, where $\Phi_0$ is the flux
quantum (in our units, $\Phi_0=2\pi$).  
In these units the counter-clockwise integral of a vector potential
$\vec A$ of the external magnetic field 
around an elementary triangular plaquette
of the grid is $2\pi f$, where $f$ is the applied magnetic flux passing through
the triangle in units of the flux quantum.  Finally, we use
$\frac{H_c^2}{4\pi} = \frac{\alpha^2}{\beta}$ as our unit of
free energy density.  Note that our units of order parameter
and free energy density are temperature-dependent.  In these units
and in the thin-wire limit without thermal fluctuations, 
the GL free energy is simply
\begin{eqnarray}
F & = & wd \int ds [ -|\psi|^2 + \frac{|\psi|^4}{2}
+ \xi_0^2 |(\nabla - iA) \psi|^2 ]~,
\label{eq:1dim-GL-F}
\end{eqnarray}
where now the integral, the vector potential,
 and the gradient are one-dimensional, running along all
wires, and $\xi_0 = \frac{\hbar}{a\sqrt{2m^*|\alpha|}}$ is the 
mean-field coherence length in zero magnetic field, in units of
the lattice spacing $a$.  We define the amplitude for the divergence of
this zero-field coherence length at $T_c^m$ via $\xi_0(t) \approx \tilde \xi_0(1-t)^{-1/2}$,
where $t=T/T_c^m$.

What order parameter patterns minimize our free energy (\ref{eq:1dim-GL-F}) for the case
of interest $f=1/2$?  Let us separate the order parameter into its
magnitude and phase: $\psi = |\psi|e^{i\phi}$.  The quantities that
are gauge-invariant and thus physical are $|\psi|$ and the gauge-invariant
phase gradient $\nabla_g\phi = (\nabla\phi - A)$, which is
proportional to the velocity of the supercurrent.  Because $\psi$ is
single-valued, the counter-clockwise
integral of $\nabla_g\phi$ around any elementary
triangle must be $2\pi(n-f)$ with $n$ an integer.  For $f=1/2$, the
free energy of a given triangle is minimized when $n-f = \pm 1/2$.
These two degenerate states consist of supercurrents flowing either
clockwise $(-)$ or counterclockwise $(+)$ around the
triangle.  In the lowest free
energy states both $|\psi|$ and $\nabla_g\phi$ are uniform around the
triangle, taking on the values $\nabla_g\phi_0 = \pm \pi/3$ and
$|\psi_0|^2=1-(\xi_0\pi/3a)^2$, provided the temperature is
low enough that $\xi_0 \leq 3a/\pi$.  For higher
temperatures close to $T_c^m$ where $\xi_0 > 3a/\pi$
the system is in the normal state for $f=1/2$.
The mean-field transition temperature $T_c^m(f)$ is minimized
at $f=1/2$ where the frustration is maximal,\cite{yn} with
$1-(T_c^m(f=1/2)/T_c^m) \approx (\tilde{\xi}_0\pi/3a)^2$ for small
$\tilde{\xi}_0/a$.  We consider the case of small $\tilde{\xi}_0/a$,
as in the recent experiments\cite{higgins}.
Then the suppression of the transition by the frustration is
small, so Ginzburg-Landau theory remains a good approximation
at $T^m_c(f=1/2)$.
To form an allowed current pattern of
the entire grid out of these local ground states of the individual
triangles one must also obey the constraint that $\psi$ remains
single-valued on passing around each elementary hexagonal plaquette.
The area of a hexagon is six times that of a triangle so $6f$ flux
quanta pass through each hexagon.  For $f=1/2$ this means the hexagons
are not frustrated and the total gauge-invariant phase change around
a hexagon must be an integer multiple of $2\pi$.  In the ground states
of the triangles the gauge-invariant phase change along each wire segment
is $\pm \pi/3$.  Thus in an allowed minimum free energy state each hexagon
either (a) has all the currents on its six edges going in the same direction
(clockwise or counterclockwise) so the total change is $\pm2\pi$
or (b) it has three edges with clockwise-going
currents and three with counter-clockwise-going currents so the
total vanishes.  The entropy
of such allowed states is roughly $0.4k_B$ per hexagonal
unit cell.\cite{baxter}

\begin{figure}
\begin{center}
\leavevmode
\epsfysize=7cm
\epsfbox{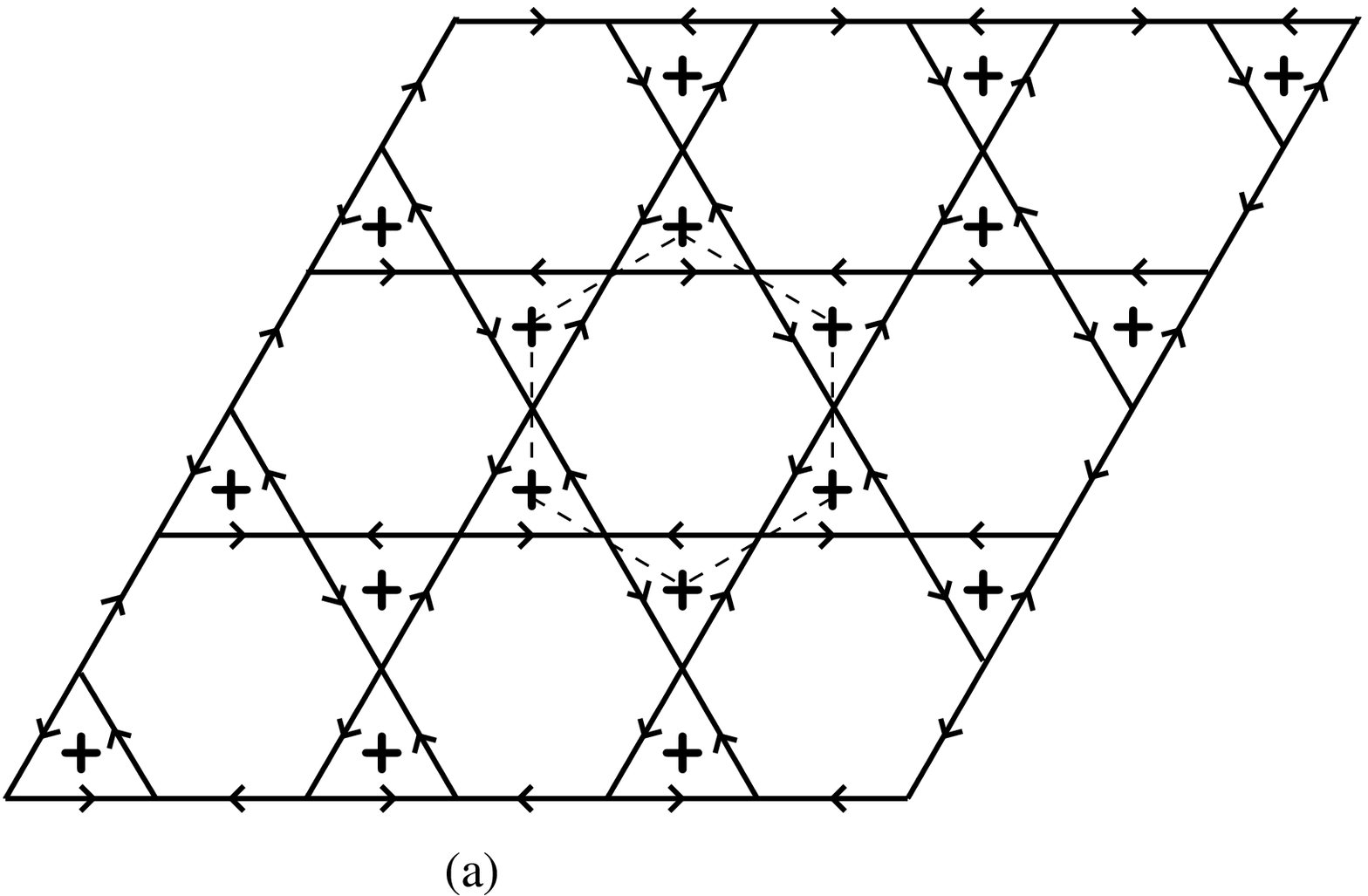}
\vspace{.3in}
\epsfysize=7cm
\epsfbox{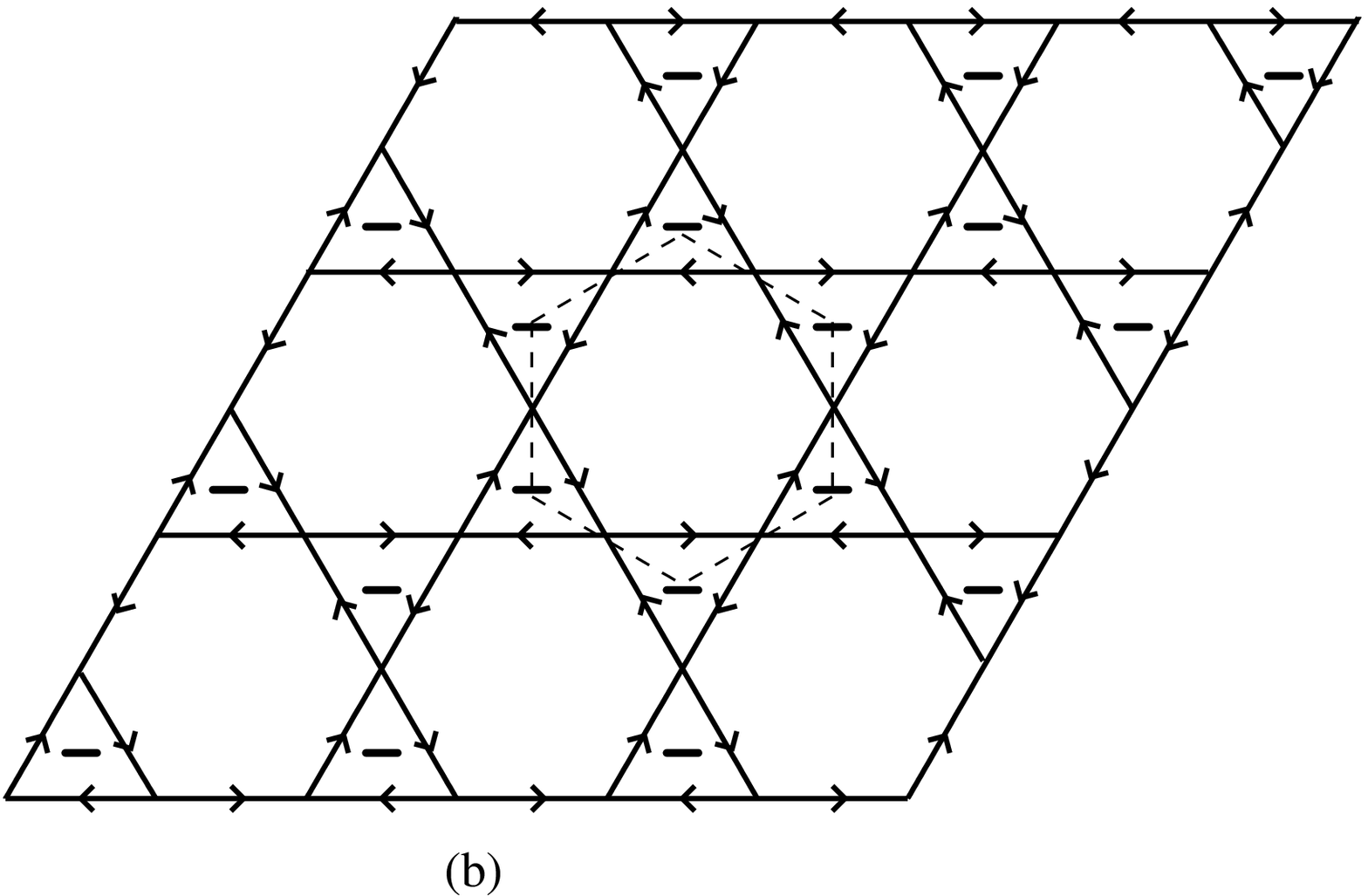}
\end{center}
\end{figure}

\begin{figure}
\begin{center}
\leavevmode
\epsfysize=6cm
\epsfbox{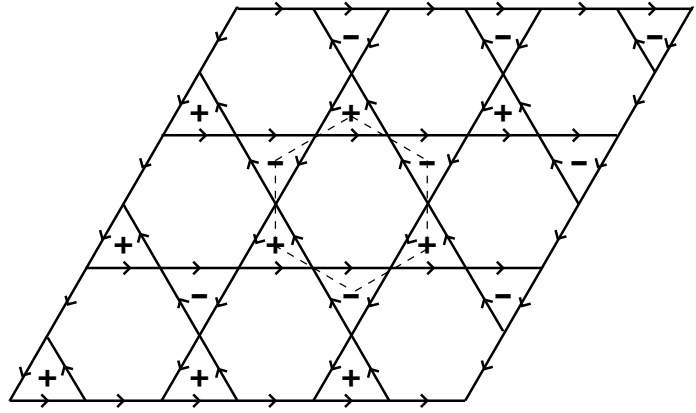}
\vspace{.3in}
\epsfysize=7cm
\epsfbox{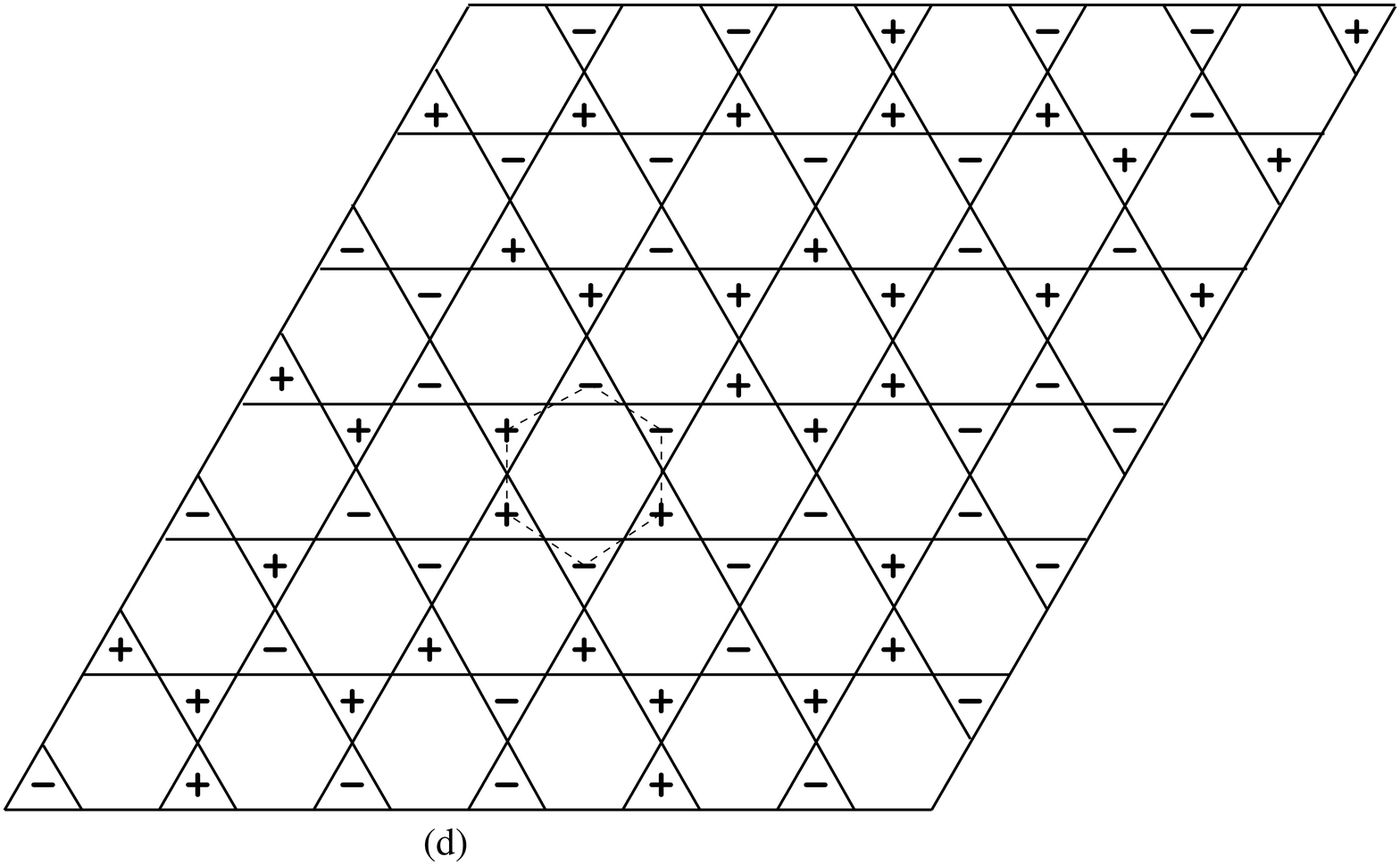}
\caption{Four minimum energy patterns of a kagome
superconducting wire network at $f=1/2$ in the thin-wire limit. 
Periodic boundary conditions are applied to the lattices. 
The arrows represent directions of the supercurrents and
$+$ ($-$) means counterclockwise (clockwise) going supercurrents
around triangles. The dotted lines denote unit cell hexagons. 
(a) the $q=0$ pattern with $(+)$ triangles only.
(b) the $q=0$ pattern with $(-)$ triangles only.
(c) one of the two $\sqrt{3}\times\sqrt{3}$ patterns: the other
pattern can be obtained by reversing the directions of all the currents.
(d) one of the many disordered current patterns.}
\label{kagome-patterns}
\end{center}
\end{figure}

Some examples of current patterns that minimize our free energy (\ref{eq:1dim-GL-F})
are shown in Fig \ref{kagome-patterns}.  
The precise degeneracy of the
ground states of (\ref{eq:1dim-GL-F}) can be viewed as occurring because both (i) each
triangle by itself has two degenerate ground states and (ii) there is
no interaction energy coupling the directions of the current circulations
on different triangles other than the constraint imposed around each hexagon.
Effects that we will consider that lift this precise degeneracy are local,
so primarily what they do is (i) lift the degeneracy of a single
triangle and/or (ii) introduce an interaction between adjacent triangles.
Case (i) favors one of the two $q=0$ patterns 
(Fig. \ref{kagome-patterns} (a), (b)), where the current
circulation has the same sense around every triangle (and necessarily
the opposite sense around every hexagon).  The $q=0$ patterns are
also selected by an interaction that favors having the same circulation
on adjacent triangles.  
An interaction 
that favors opposite circulations on adjacent triangles selects the
\sqtt\ patterns (Fig. \ref{kagome-patterns} (c)).
The majority of the minimum-energy current patterns
are disordered; we call them
``random'' patterns. One of the ``random'' patterns is shown in
Fig. \ref{kagome-patterns} (d).  There also exist
periodic patterns other than the 
$q=0$ and $\sqrt{3}\times\sqrt{3}$ patterns.
But to stabilize these other periodic patterns, 
we would need further-neighbor
interactions between triangles to be comparable to the single-triangle
or nearest-neighbor triangle interactions, which seems unlikely to occur.

If the degeneracy of all these current patterns is not lifted strongly
enough, the system will not order into one pattern.  Instead it will,
by phase-slip processes, move among many patterns, with the
equilibrium probability of each pattern being given by its Boltzmann factor. 
Let us consider the effects of these configurational fluctuations
among ground-state current patterns of (\ref{eq:1dim-GL-F}), first neglecting
other types of thermal fluctuations of $\psi$ such as vortices or
phase fluctuations.  Let us consider the phase correlations.  If our system
does order into one particular current 
pattern (the most likely candidates being
the $\sqrt{3}\times\sqrt{3}$ and $q=0$ patterns) then if the phase is known
at one point on the grid, it is known everywhere.  This is the usual
situation in a superconducting phase.  (Of course, since we are in two
dimensions (2d), thermal fluctuations of the
phase will reduce this long-range order
to power-law quasi-long-range order of $\psi$ and vortices will destroy
the superconductivity above a Kosterlitz-Thouless (KT) transition.)
Now consider the situation when the system is fluctuating freely among
different ground-state current patterns:  
If the phase is known at one junction of
the grid, then we know the gauge-invariant phase change to a neighboring
junction is $\pm\pi/3$.  Thus the phase there can take on two different
values differing by $2\pi/3$.  This uncertainty in the phase increases
with distance, meaning $\psi$
will not show any long-range order.  But there is no uncertainty in
$\psi^3 = |\psi|^3e^{i3\phi}$.  Thus even when the system fluctuates freely
among the different current patterns it has long-range order in $\psi^3$.
This is the novel superconducting phase, where Cooper pairs do not show
phase coherence at distances large compared to the
grid spacing, but ``Cooper sextuples'' do.
Again, since we are in 2d, thermal fluctuations of the phase reduce this to
quasi-long-range order, and (fractional) vortices unbind and destroy it at a
KT transition.
There is another possibility for a superconducting
state of this system: it could be out of equilibrium and kinetically
frozen in a particular disordered current pattern because the free energy
barriers to phase-slip that must be crossed to 
move among the different patterns are
too high compared to the temperature.  If the system is in the novel
``$\psi^3$'' phase and is cooled to low enough temperature, this freezing
into a sort of kinetic vortex glass state will certainly occur.

If the degeneracy of the current patterns is lifted strongly enough, 
then the system will be ordered into a conventional $\psi$ ordered phase,
either the $\sqrt{3}\times\sqrt{3}$ or $q=0$ pattern. For the $q=0$ pattern
to be stabilized, we will assume that
the energy must be approximately $0.4k_B T$ per unit cell
lower than most of the other patterns.  This is because the $q=0$ ordered phase 
does not have an extensive entropy of circulation patterns, while the 
allowed patterns have an entropy of roughly $0.4k_B$ per unit cell.  When all 
ground-state current patterns
are given the same weight, the system is almost ordered (i.e., has power-law
spatial correlations) in the \sqtt\ pattern, and a small degeneracy-lifting
term favoring the \sqtt\ pattern will stabilize that phase.\cite{huse}
The reason the \sqtt\ phase is stabilized so readily is that the system can
make local rearrangements of the circulation patterns without destroying the
long-range order of this phase, so this phase (unlike the $q=0$ phase)
actually has extensive entropy comparable in magnitude to that of the
$\psi^3$ phase.  We have not attempted a more precise statistical-mechanical
determination of the phase diagram as a function of the two parameters: the 
single-triangle degeneracy-lifting term and the nearest-neighbor
inter-triangle interaction.  Such a study would be of interest and would
more precisely delineate when the novel $\psi^3$ phase is stable.

Among the three possibilities, which superconducting state the system will be in 
depends on the strength of the degeneracy-lifting effects and the free energy 
barriers to move vortices at a KT temperature. Above the KT temperature,
the superconductivity will be suppressed so there is no ordered phase.
Far enough below the KT temperature, because of the large free energy barriers
to vortex motion, the system will be kinetically frozen into a particular
pattern of $\psi$ (we call this a kinetic vortex glass state) so that 
experimentally the equilibrium state is not accessible.
So in order to explore the three possible superconducting phases, in the 
following sections, we estimate three things: 
(i) the helicity modulus of each current pattern to estimate the KT transition 
temperatures of the $\psi$ ordered phases and the novel $\psi^3$ ordered phase 
within the Ginzburg-Landau thin-wire limit, 
(ii) the amount by which the degeneracy of the current patterns is lifted 
by various degeneracy-lifting effects 
(inductive couplings between adjacent supercurrents, 
non-zero wire width, wire bending, and order-by-disorder 
effect (thermal fluctuations)), and 
(iii) the temperature where phase-slip is frozen out due to large barriers 
to vortex motion.

\section{ Helicity Modulus  }

In this section, we estimate the helicity moduli of the different
minimum-energy current patterns in order to estimate their
Kosterlitz-Thouless (KT) transition temperatures.
The free energy cost of a long-wavelength phase distortion \cite{ch_lu}
is
\begin{eqnarray}
\Delta F&=& \frac{1}{2} \int \: d^2 \vec{r} \: \rho_s(T) 
(\nabla \delta\phi(\vec{r}))^2 ~,
\label{eq:def-helmod}
\end{eqnarray}
where $\delta\phi$ is the deviation of the phase of the order
parameter away from the minimum-free-energy pattern,  
and $\rho_s(T)$ is the helicity modulus of that pattern. 
The integral here is over the full two-dimensional space (not just the wires)
since the helicity modulus is defined for the continuum approximation
to the system.  The renormalization group (RG) theory 
of the KT transition \cite{kosterlitz} 
shows that for a conventional superconducting phase,
the vortices unbind and the helicity modulus 
vanishes discontinuously at a KT 
temperature $T_{KT}$, with 
\begin{eqnarray}
\lim_{T\rightarrow T^{-}_{KT}} \; \rho_s(T)&=&
\frac{2}{\pi}k_B T_{KT}~.
\label{eq:rhos_Tkt}
\end{eqnarray}  
For the $\psi^3$ phase, on the other hand,
the vortices that unbind are 1/3-vortices, so 
attract one another with a potential that is 1/9 of that for full
vortices.  Thus the KT criterion for the $\psi^3$ phase is instead
\begin{eqnarray}
\lim_{T\rightarrow T^{-}_{KT,1/3}} \; \rho_s(T)&=&
\frac{18}{\pi}k_B T_{KT,1/3}~.
\label{eq:rhos_Tkt3}
\end{eqnarray}  
We consider the helicity modulus of each minimum-energy current
pattern of the thin-wire model.  
Among all the current patterns, we choose to study the helicity moduli
of the two simple patterns: the \qz\ and \sqtt\ patterns 
(Fig. \ref{kagome-patterns} (a),(b),(c)).  These appear to be the extremal
cases, with the helicity moduli of other patterns being intermediate
between these two.
Our analysis shows that the \sqtt\ pattern has a smaller helicity modulus
than the \qz\ pattern, and that the helicity moduli of all current patterns
become identical in the London limit of $\xi_0 \ll a$.  

When a long-wavelength distortion of the phase 
of the order parameter is applied to
a particular minimum-energy current pattern, within Ginzburg-Landau theory
we have to take into account the
distortion of the magnitude of the order parameter as well as that of 
the phase.  In the London limit these magnitude distortions are negligible.
But we find that near the mean-field transition temperature, the effect on the
helicity modulus due to magnitude distortions is substantial. 
If we write the magnitude distortion as $\delta|\psi|$ and
the phase distortion as $\delta\phi$, the distortion of the complex
order parameter $\delta\psi$ will be, up to first order in $\delta|\psi|$
and $\delta\phi$, 
\begin{eqnarray}
\delta\psi&=& (\delta |\psi| + \imath |\psi_0| \delta \phi) e^{\imath\phi_0} ~,
\label{eq:complex-fluct}
\end{eqnarray}
where $|\psi_0|$ and $\phi_0$ represent the minimum energy magnitude and phase
of the order parameter in the particular $\psi$ pattern being considered.  
The difference between the (thin-wire) GL free energies 
with and without the distortion is, up to second order 
in $\delta |\psi|$ and $\delta \phi$,
\begin{eqnarray}
F[\psi_0+\delta\psi]-F[\psi_0]&=& wd \; \int ds
\; [ \; 2|\psi_0|^2 (\delta |\psi|)^2 + \xi_0^2 (\nabla \delta |\psi|)^2 
\nonumber \\
& &+ \xi_0^2 |\psi_0|^2 (\nabla \delta \phi)^2  
+ 4 \xi_0^2 |\psi_0| (\nabla_g \phi_0 \cdot \nabla \delta \phi) \delta |\psi|
 \; ]~, \label{eq:dev-energy}
\end{eqnarray}
where $|\nabla_g \phi_0|=\pi/3$.  
The term $ (\nabla_g \phi_0 \cdot \nabla \delta \phi) \delta |\psi|$
couples amplitude and phase and will be negative to minimize the free
energy cost of any spatially nonuniform phase distortion.
Without the magnitude distortion (i.e. for $\delta|\psi| = 0$), 
one can see from (\ref{eq:dev-energy}) that the 
helicity modulus does not depend on the  
current pattern; this is the result in the London limit. 

Minimizing the free energy difference with respect to the order parameter
magnitude and phase, we obtain the (Euler-Lagrange) equations that apply
along each wire segment: 
\begin{eqnarray}
2 |\psi_0|^2 \delta |\psi| + 2 \xi_0^2 |\psi_0| \nabla_g
\phi_0 \cdot \nabla \delta \phi -  \xi_0^2 \nabla^2 \delta|\psi| &=& 0~,
\label{eq:dfdpsi}\\
2  \nabla_g \phi_0 \cdot \nabla \delta |\psi| + 
 |\psi_0| \nabla^2 \delta \phi = 0~. \label{eq:dfdphi}
\end{eqnarray}
For the current patterns of interest,
these equations are homogeneous and linear, so are
straightforwardly solved for each wire segment.
The boundary conditions at the junctions are that
$\delta|\psi|$ and $\delta\phi$ are continuous
and that the sums of the first derivatives of $\delta|\psi|$ and $\delta\phi$ 
moving away from each junction $i$ along each of the 4 wire
segments meeting at that junction must vanish:
\begin{eqnarray}
\sum_{j=1}^{4} \nabla_{ij} \delta |\psi||_i = 0~, 
& &\sum_{j=1}^{4} \nabla_{ij} \delta \phi|_i = 0~, \label{eq:bdc}
\end{eqnarray}
where $\nabla_{ij}$ denote the derivative taken with the position
coordinate increasing as one moves from junction $i$ to neighboring
junction $j$.
The supercurrents are conserved if all the above 
conditions are satisfied.  

\begin{figure}
\begin{center}
\leavevmode            
\epsfysize=5cm
\epsfbox{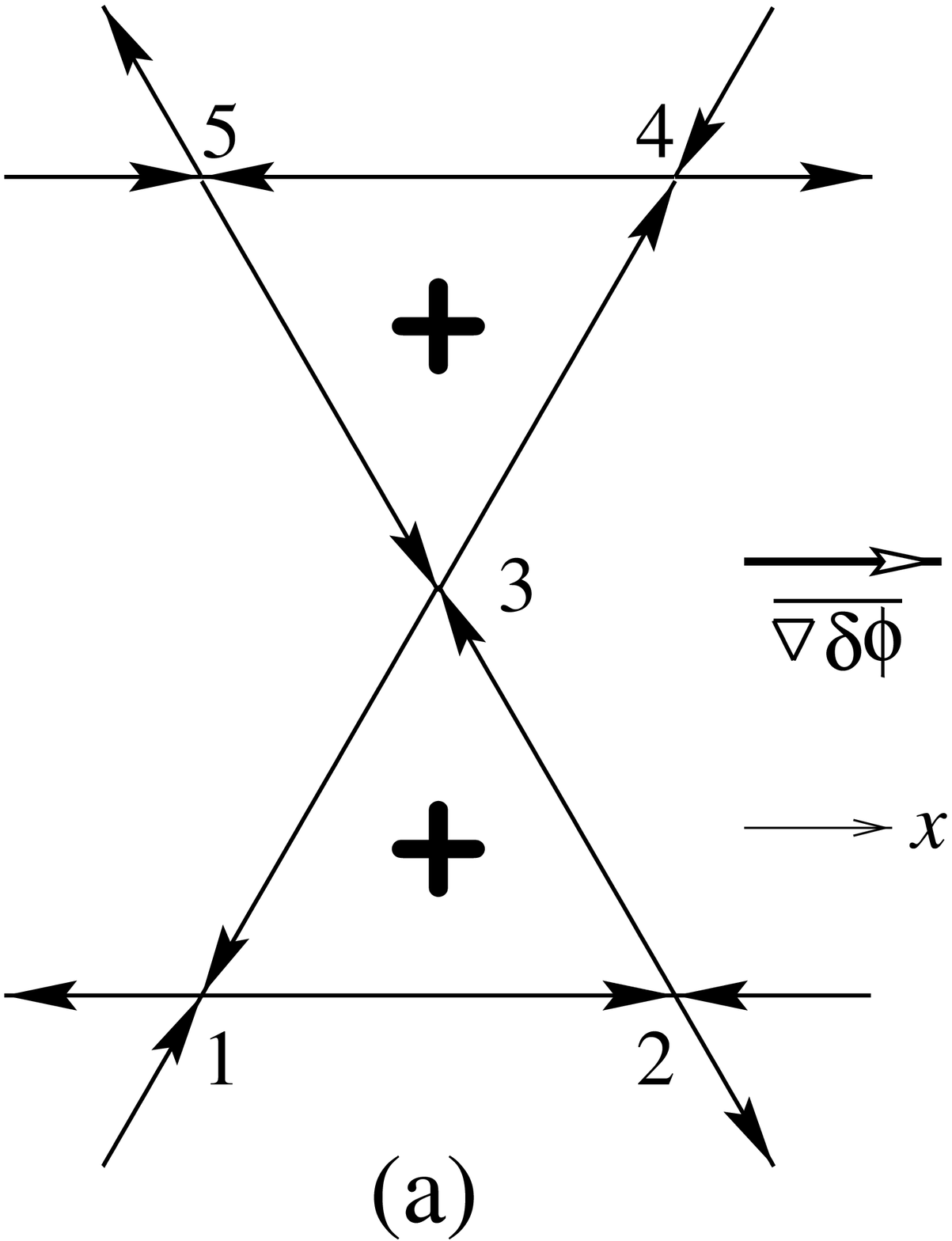}
\hspace{.7in}
\epsfysize=5cm
\epsfbox{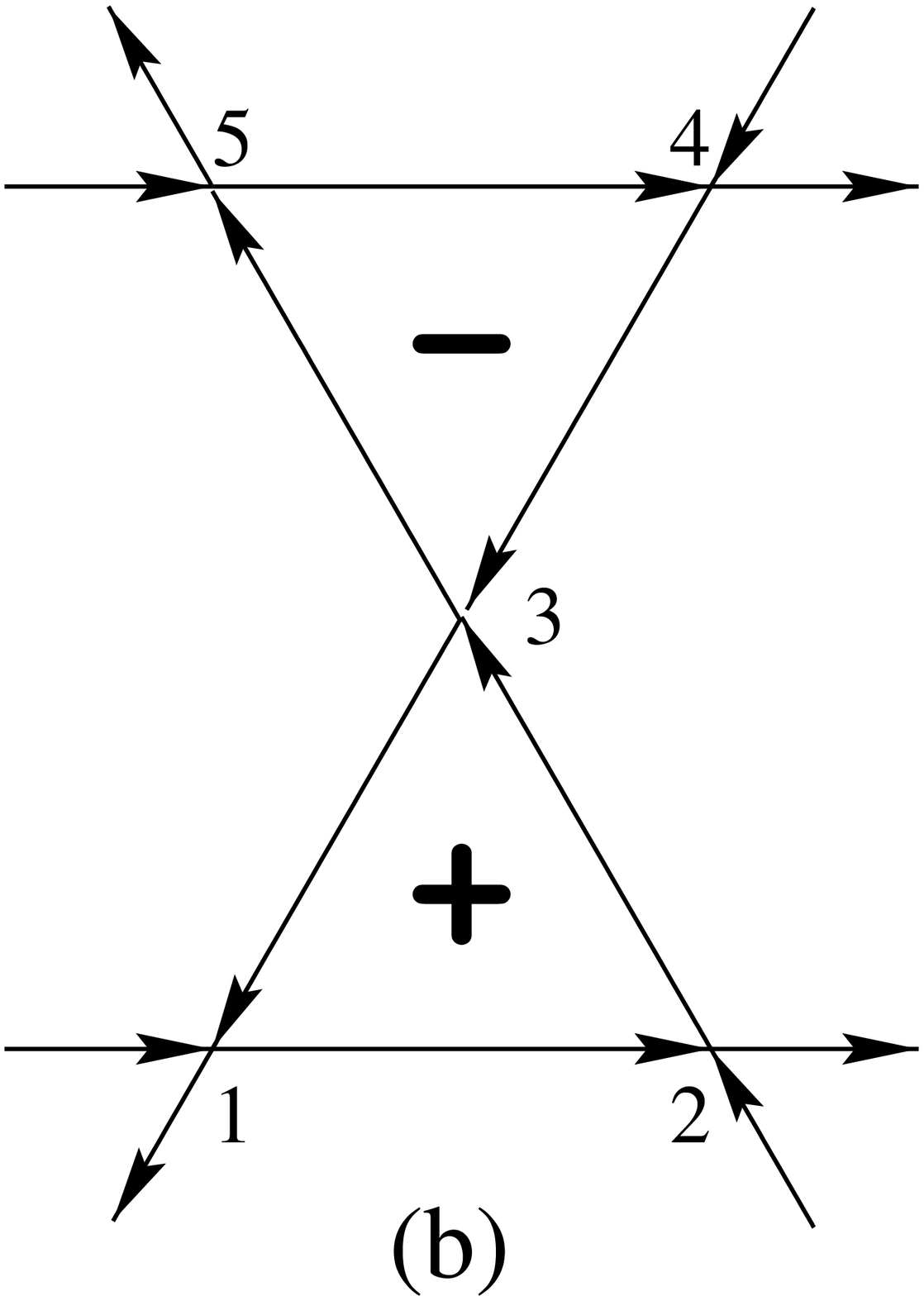}
\caption{Gauge-invariant unit cells of the \qz\ (a) and \sqtt\ pattern (b). 
Arrows represent
the supercurrents. The junctions are marked by numbers.
The small large-scale average phase gradient $\overline{{\vec \nabla} \delta \phi}$ 
added to measure the helicity moduli is along the
$x$ direction.}
\label{hourglasses}
\end{center}
\end{figure}

Fig. \ref{hourglasses} shows the gauge-invariant 
unit cells of the two current patterns
that we examine.  For the purpose of the present discussion, we choose
the unit cell to consist of two adjacent triangles (elsewhere we use
a single hexagon as an alternative unit cell).  To calculate the
helicity modulus, we apply a small added phase gradient that is uniform
at scales larger than a unit cell, with a spatial average of
$\overline{{\vec \nabla} \delta \phi}$.  With this gradient the new
minimum free energy state has a pattern of the
gauge-invariant variables $\delta|\psi|$ and $\nabla \delta\phi$ that
is the same in every unit cell, but $\delta\phi$ changes systematically
between cells:
\begin{eqnarray}
\delta |\psi|(\vec{r}+\vec{R}) & = & \delta |\psi|(\vec{r})~, \label{eq:bloch-1-Q0} \\
\nabla \delta \phi(\vec{r}+\vec{R}) & = & 
\nabla \delta \phi(\vec{r})~, \label{eq:bloch-2-Q0} \\
\delta \phi(\vec{r}+\vec{R}) & = & \delta \phi(\vec{r}) +
\vec{R} \cdot \overline{{\vec \nabla} \delta \phi}~,
\end{eqnarray}
where $\vec{R}$ is any lattice translation vector connecting equivalent
points in two different unit cells.  Along each wire segment, the added
phase gradient $\nabla \delta \phi$ is either of the same or opposite
sign from the gauge-invariant phase gradient $\nabla_g \phi_0$ that is
already present.  This results in a change in the magnitude of the
supercurrent, and a resulting decrease or increase, respectively, in
$|\psi|$.
For our two patterns the helicity modulus is isotropic.  Let us put the
added phase gradient along the $x$ direction, as illustrated 
in Fig. \ref{hourglasses}.  Let us look at what happens in the
vicinity of junction 3 in Fig. \ref{hourglasses}.
For the \qz\ pattern (Fig. \ref{hourglasses} (a)), the added phase
gradient increases the supercurrent in two of the adjoining
wires, while it decreases it in the other two.  The net effect is
that, to lowest order in the added phase gradient, $\delta|\psi|=0$
at the junction.  This is true at all the junctions for
the \qz\ pattern for any orientation of the added phase gradient.
For the \sqtt\ pattern (Fig. \ref{hourglasses} (b)), on the other hand,
the added phase gradient decreases the supercurrent on all the wires
adjacent to junction 3, and therefore increases $|\psi|$ at
that junction.  A decrease in $|\psi|$ also occurs at the other two
junctions.  This additional ``relaxation'' of $|\psi|$ at the junctions
is connected with a lower free energy cost of the added phase
gradient and thus a lower helicity modulus in the \sqtt\ pattern 
than in the \qz\ pattern. 
 

From the definition of the helicity modulus (\ref{eq:def-helmod}), 
the energy cost per unit cell due to the added phase distortion is
\begin{eqnarray}
 \Delta F &=&  A_{hex} \; 
\frac{1}{2} \rho_s(T)  (\overline{{\vec \nabla} \delta \phi})^2~,
\label{eq:Delta-F}
\end{eqnarray}
where $A_{hex}=2\sqrt{3}a^2$ is the area of a unit cell.
We have obtained the helicity moduli of both patterns at all
temperatures.  The results simplify near the mean-field phase
transition where,
surprisingly, the helicity modulus of the \sqtt\ pattern, $\rho_s$[\sqtt],
vanishes much faster than that of the \qz\ pattern, 
$\rho_s$[\qz], as $|\psi_0| \rightarrow 0$. 
For $|\psi_0| \ll 1$ we obtain
\begin{eqnarray}
\rho_s [\qz]& \approx &  |\psi_0|^2 \; E_c/\pi~,
\label{eq:helmod-q0} \\
\rho_s [\sqtt]& \approx & \frac{2}{3\sqrt{3}}|\psi_0|^4 \; E_c~,
\label{eq:helmod-sqrt33} 
\end{eqnarray}
where
\begin{eqnarray} 
E_c&\equiv&\frac{3H_c^{2}}{4\pi} \; w \cdot d \cdot a~ 
\label{eq:E_cond} 
\end{eqnarray}
is the condensation energy per unit cell in zero field.  Note that $E_c$
remains nonzero at the $f=1/2$ transition temperature.
At low temperature in the London regime where $\xi_0 \ll a$, 
the magnitude distortions are negligible and the the helicity moduli of 
all of the current patterns are $\rho_s \approx 2(\xi_0/a)^2E_c/\sqrt{3}$. 
More generally, the helicity moduli are
\begin{eqnarray}
\rho_s [\sqtt]&=&\frac{E_c}{2\sqrt{3}} \; 
\frac{\xi_0^4 \eta^2 (\frac{2 \cos{\eta} + 1}{\sin{\eta}}) }
{\frac{6 (\nabla_g \phi_0)^2}{|\psi_0|^2 \eta} \xi_0^2
-(\frac{2 \cos{\eta} + 1}{\sin{\eta}}) },
\label{eq:red-kt33} \\
\rho_s [\qz]&=&\frac{E_c}{2\sqrt{3}} \;  \frac{\xi_0^4 \eta^2}
{\frac{4 (\nabla_g \phi_0)^2}{|\psi_0|^2 \eta} \xi_0^2
(\frac{1-\cos{\eta}}{\sin{\eta}}) - 1}, \label{eq:red-ktq0} \\
\frac{\eta^2}{2} &\equiv&   3 (\nabla_g \phi_0)^2 - \xi_0^{-2} , \nonumber 
\end{eqnarray}
where the temperature dependence is in $\xi_0$, $|\psi_0|$, $\eta$, and $E_c$.

These helicity moduli can be used to estimate the Kosterlitz-Thouless (KT)
transition temperatures of the various possible phases.  Putting in the
experimental parameters from Appendix A we obtain for the conventional
$\psi$ ordered phases
$1-\frac{T_{KT,\sqrt{3}\times \sqrt{3}}(f=\frac{1}{2})}{T_c^m(f=\frac{1}{2})}
\approx0.0017$ and $1-\frac{T_{KT,q=0}(f=\frac{1}{2})}{T_c^m(f=\frac{1}{2})}
\approx 0.0007$. 
Note that the KT transition temperature is lower for the \sqtt\ phase
because of its smaller helicity modulus.  Stability of the novel
$\psi^3$ phase requires a nine time larger helicity modulus, so this phase's
KT transition occurs at lower temperature.  For our parameters it is in the
crossover between GL and London regimes, where the difference in helicity
moduli between patterns is not very large.  Because 
the random current patterns are locally more similar to the \sqtt\ patterns
than they are to the \qz\ patterns, we will use the helicity modulus of
the \sqtt\ pattern (\ref{eq:red-kt33}) to estimate the KT transition 
of the $\psi^3$ phase.  This leads to
$1-\frac{T_{KT,1/3}(f=\frac{1}{2})}{T_c^m(f=\frac{1}{2})}\approx 0.007$.


\section{Degeneracy-Lifting Effects}

Now that we have estimated the KT temperatures of the $\psi$ ordered
phases and the $\psi^3$ ordered phase, in this section we consider 
effects that lift the degeneracy of the minimum energy 
patterns at $f=1/2$, estimating 
the strength of each effect at the KT temperatures.
The net result of these effects determines which of the possible
superconducting phases occur in this system.
First we consider
the inductive coupling between the supercurrents in
different triangles, finding that this effect is negligible 
near the KT transitions for the experimental parameters we 
consider.  The effects that can substantially lift the degeneracy
are the nonzero width of the wires, a possible bending of the wires
away from straight, and ``order-by-disorder'' effects due to thermal
fluctuations in the Ginzburg-Landau regime.

\subsection{Magnetic Energy}

One effect that lifts the degeneracy of the ground-state current patterns in
the thin-wire limit of mean-field theory is the magnetic energy due
to the inductance of the grid and the supercurrents.  Here we
estimate this effect, finding that it is quite small at the KT transition
for parameters corresponding to the recent experiments \cite{yi}\cite{higgins}.
Since the magnitudes
of the currents flowing in every triangle are identical, the energy due
to self-inductance of each triangle is the same for every current pattern.
The energy due to mutual inductance \cite{jackson} is
\begin{eqnarray}
F_m &=& {1 \over 2}\sum_{i \neq j} M_{ij}I_{i}I_{j}~, \\
M_{i,j}&=&\frac{1}{c^2}\oint_{C_i} \oint_{C_j} 
\frac{d \vec{l}_i \cdot
d \vec{l}_j}{|\vec{r}_i -\vec{r}_j|}~,
\end{eqnarray}
where $M_{ij}$ is the mutual inductance between triangles $i$ and $j$
and $I_i$ is the supercurrent in triangle $i$. 
$C_i$ is the closed contour around the $i$th triangle.
At large distance $r$, the interaction is a magnetic dipole-dipole
coupling, so falls off as $r^{-3}$, making the sum well-convergent.
In fact, the interaction between nearest-neighbor triangles is
dominant in the sum.
If only nearest-neighbor pairs of triangles are included,
$F_m(\sqtt)-F_m(\qz) \cong 0.53\frac{a I^2}{c^2}$.
(The second and third neighbor contributions are an order of magnitude smaller at
$0.052\frac{a I^2}{c^2}$ and $0.032\frac{a I^2}{c^2}$,
respectively.) Thus the \qz\ pattern has the lowest magnetic energy.

The magnetic energy difference
per unit cell for our experimental parameters
(Appendix A) is approximately $50 k_B T_c 
(1-\frac{T}{T_c^m(f=1/2)})^2$. Since 
$(1-\frac{T_{KT,1/3}(f=1/2)}{T_c^m(f=1/2)})$
is about $0.007$, the magnetic energy difference
is about $0.002 k_B T_c$ at $T_{KT,1/3}$, which is
negligible compared to the other degeneracy-lifting effects
that we examine below.

\subsection{Finite Wire Width}

So far, we have assumed that the wires can be
considered as perfect mathematical lines and the junctions 
as perfect points.
In real experiments, the wires have a non-zero width $w$, and the supercurrent
density is nonuniform across the width of the wires and at the junctions.
In this section, we estimate how much this non-zero 
wire width lifts the degeneracy of the minimum energy current patterns.
The degeneracy of the two
ground states of a single triangle is lifted, favoring the
$q=0(--)$ pattern. Here we assume that the wires are 
straight, as in the published experiment \cite{higgins}. 

As a simple model for a rough description and estimate of this effect,
let us approximate a single triangle as a circle, 
and ignore its connections to the other triangles. 
Below we treat the kagome grid numerically to check when this crude
model is qualitatively correct. The model we use is a circular wire loop of width
$w$, occupying the area between concentric circles of radii $1-w/2$
and $1+w/2$. The uniform applied field is such that one-half of a flux
quantum passes through the circle whose perimeter runs along the center
of the wire. The two nearly-degenerate current patterns both have supercurrents
flowing only around the loop, so the gauge-invariant phase gradient 
${\vec \nabla_g \phi}$ has no radial component; the currents flow in opposite
directions in the two patterns. The $(-)$ pattern
has no vortex within the loop, and has
\begin{eqnarray}
|\nabla_g \phi_{-}| &=& \frac{r}{2}~,
\label{eq:fww_m}
\end{eqnarray}
at radius $r$. The $(+)$ pattern has a vortex within the loop, resulting in
\begin{eqnarray}
|\nabla_g \phi_{+}| &=& \frac{1}{r}-\frac{r}{2} ~.
\label{eq:fww_p}
\end{eqnarray}
The energies $E_L^{(\pm)}$ of these supercurrents are readily obtained
in the London approximation.
The $(-)$ pattern has the lower energy.  For $w\ll a$ the energy difference is
of order $(w/a)^2 E_L$.
We verify below that this simple argument
provides the correct sign and order in $(w/a)$ of
the energy difference, but the coefficient of the leading term 
is not estimated accurately 
with this rough model.
This is because in the proper kagome geometry the details of the current patterns
around the junctions
that have been ignored in this crude model contribute 
substantially to the small energy 
difference.  The essential conclusion here is that the lowest energy pattern
has the vortices in the hexagons and not in the triangles.

\begin{figure}
\begin{center}
\leavevmode
\epsfysize=6.5cm
\epsfbox{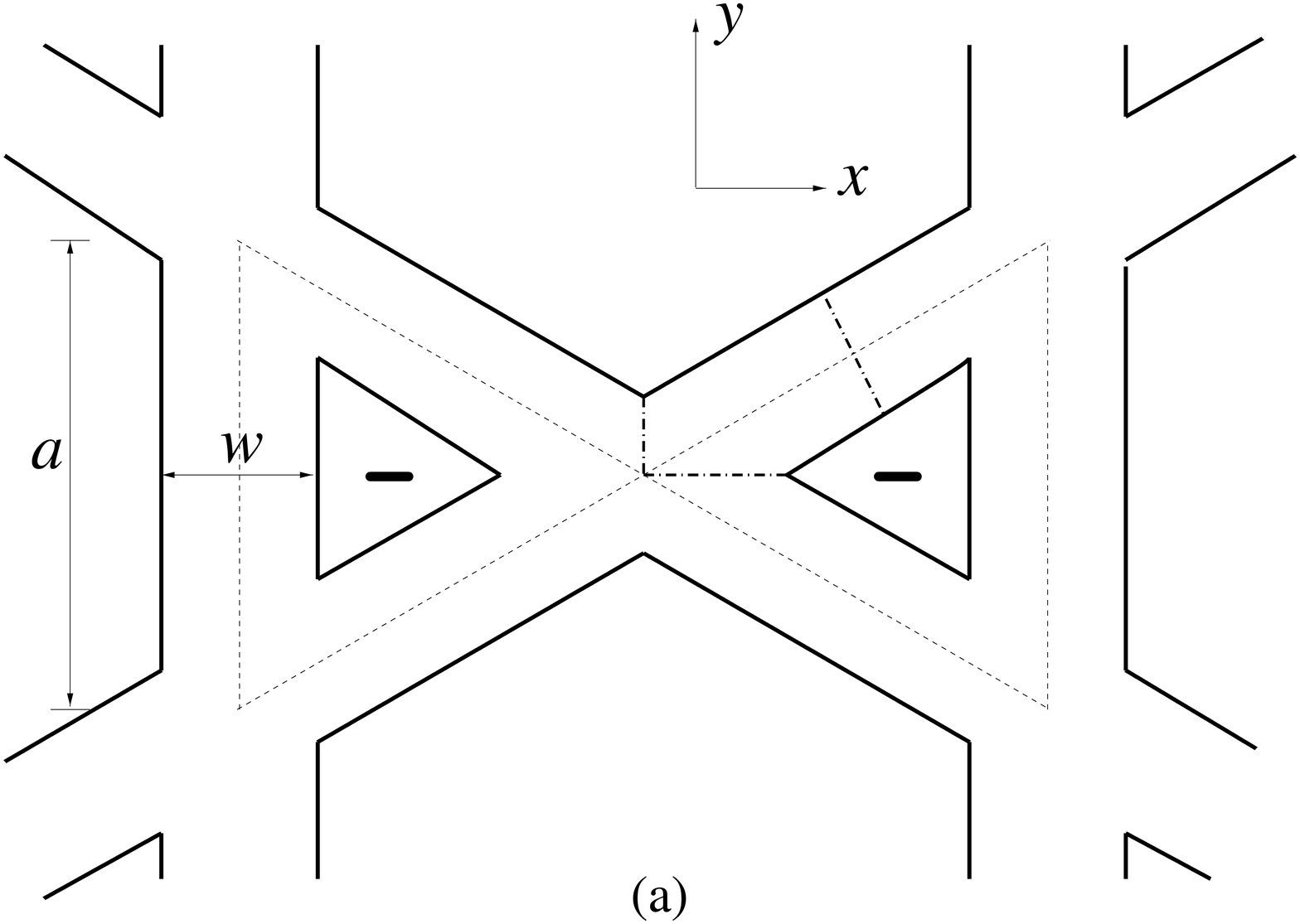}
\vspace{.7in}
\epsfysize=6.5cm
\epsfbox{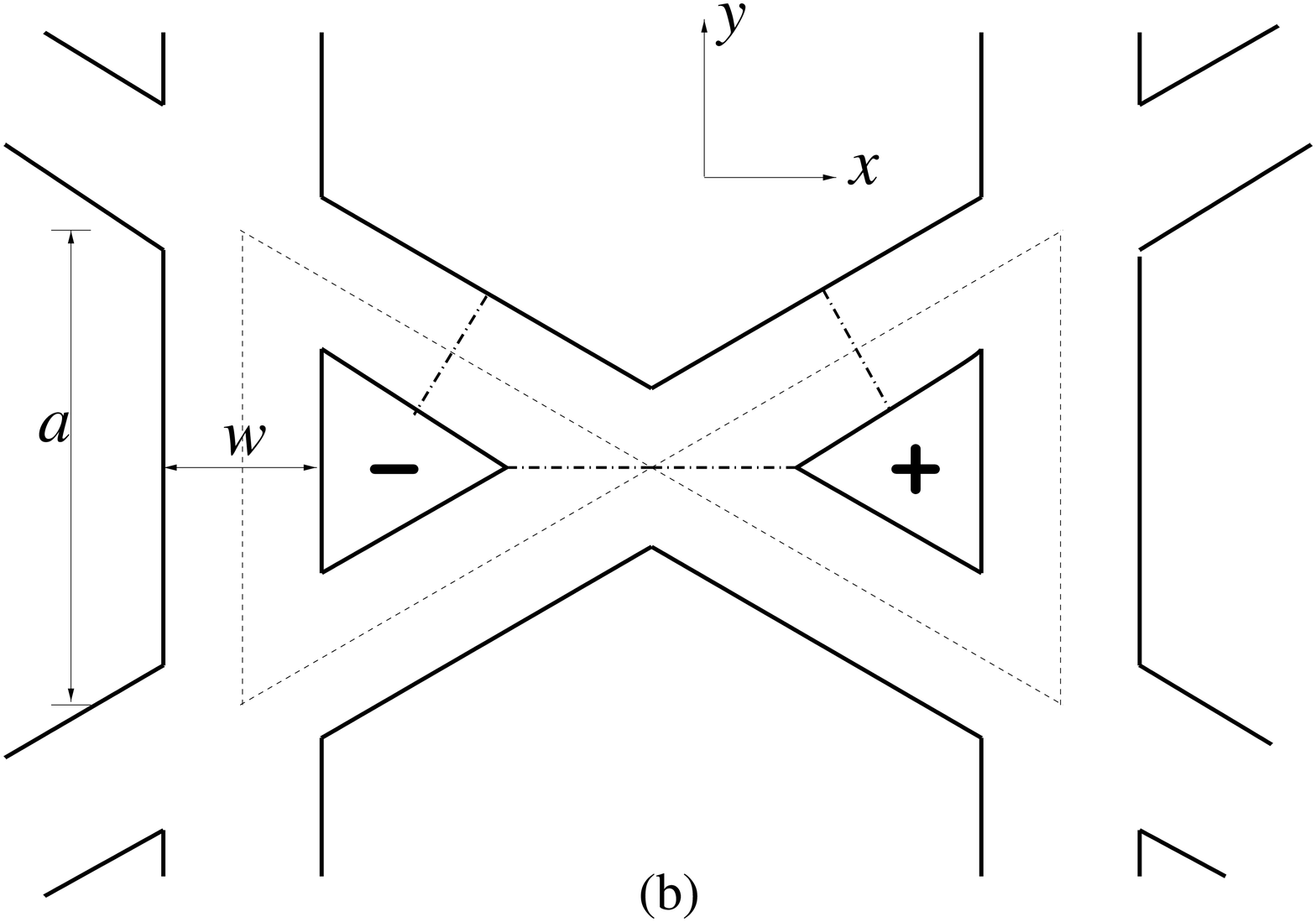}
\caption{Part of the kagome-lattice wire network with
non-zero wire width $w$. (For clarity, the wire width is shown larger than
the values $w/a \approx 0.1$ examined in the recent experiments.)  A half 
of one flux quantum penetrates through each of the triangles surrounded 
by the dotted lines along the centers of the wires. 
Because of the symmetries of the order parameter and supercurrent patterns, 
we need to solve for the patterns only in the areas 
confined by the dash-dotted lines and the boundaries of the wires.
(a) In the $q=0$ patterns, only one twelfth of a unit cell is needed. 
(b) In the $\sqrt{3}\times \sqrt{3}$ pattern, only one sixth of a unit cell is needed.
}
\label{finite-ww}
\end{center}
\end{figure}

We can go one step further with this simple model, and estimate the
energies in Ginzburg-Landau theory instead of the London approximation.
For $w\ll a$ the relative variation of the order parameter magnitude
across the wires is only of order $(w/a)^3$, and the resulting
Ginzburg-Landau contribution to the energy difference is of order
$(w/a)^4E_L$, and thus only a small correction to the result from
the London approximation for the relatively
narrow wires we are interested in.  We have numerically obtained the
minimum free energy order parameter patterns within Ginzburg-Landau
theory on the proper kagome lattice to verify that we are indeed justified
in using the simpler London approximation~\cite{thesis}.

The London energy that we consider is
\begin{eqnarray}
F_L & = & d \; \xi_0^2 |\psi_0|^2 \; \int d{\cal A}~(\vec{\nabla}\phi - \vec{A})^2 ~,
\label{eq:London-E-num}
\end{eqnarray}
where the integral runs over the area occupied by the wires,
as illustrated in Fig. \ref{finite-ww}.  We obtain the London energy of
the \sqtt\ pattern and of the the two different $q=0$ patterns.
Each of these current patterns has a high symmetry, so we need to
solve for the pattern only in a small fraction of the unit cell,
as indicated in Fig. \ref{finite-ww}.
We discretize the system with an equilateral triangular numerical grid where
nearest-neighbor grid vectors are aligned parallel to the edges of the wires.
We represent the patterns in terms of gauge-invariant phase differences only.
The uniform magnetic field induces net diamagnetic currents circulating around
each elementary plaquette of our numerical grid.  In each pattern 
the gauge-invariant phase difference between the centers of adjacent wire
junctions along the center line of the connecting wire is fixed to be $\pm\pi/3$ 
depending on the direction of the supercurrents. 
For each grid spacing we use, we put as many grid points as 
possible along all boundaries.

For each pattern we start with physically-sensible smooth initial conditions 
that satisfy the 
above constraints.  Then we minimize the discretized version of 
the London energy for the \qzp,\ \qzm,\ and \sqtt\ 
patterns using the relaxation method.  We have done this for various values
of the normalized wire width, $w/a$. 
For each wire width, we use various different grid spacings.
We then use these results for different grid spacings to extrapolate to 
the continuum limit.~\cite{thesis}

\begin{table}
\begin{center}
\caption{London energy and energy differences between the different current
patterns in units of 
$E_L \equiv 2(\frac{\xi_0}{a})^2 |\psi_0|^2 (\pi/3)^2 E_c$
for various wire widths. $F_{++}$, $F_{--}$, and $F_{-+}$
denote the London energy per unit cell for the \qzp,\ \qzm,\ 
and \sqtt\ patterns, respectively. }
\label{table:London-1}
\vspace{.3cm}
\begin{tabular}{l|l|l|l|l|l|l} \hline \hline
$w/a$ & $F_{++}$ & $F_{--}$ & $F_{-+}$ &
$F_{++}-F_{--}$ & $F_{++}-F_{-+}$ & $F_{-+}-F_{--}$
\\ \hline
0.0201 & 1.02344 & 1.02091 & 1.02049 & 0.00253 & 0.00295 &
-0.00042 \\
0.0422 & 1.05662 & 1.04680 & 1.04831 & 0.00982 & 0.00831 &
0.00151 \\
0.0619 & 1.09615  & 1.07465 & 1.08018 & 0.0215 & 0.01597  &
0.00553  \\
0.0825 & 1.14494 & 1.10582 & 1.11804 & 0.03912 & 0.0269 &
0.01222 \\
0.1019 & 1.19840 & 1.13736 & 1.15836 & 0.06104 & 0.04004 &
0.021 \\
0.1237 & 1.26741 & 1.17495 & 1.20899 & 0.09246 & 0.05842   &
0.03404 \\ \hline \hline
\end{tabular}
\end{center}
\end{table}

Our results for this non-zero wire width effect are presented in 
Table \ref{table:London-1}.  Except for extremely thin wires,
the \qzm\ pattern has the lowest energy, $F_{--}$,
and the energy of the \qzp\ pattern, $F_{++}$, is higher by
of order $(w/a)^2 E_L$, consistent with our crude ``circle'' model.
If all that were happening is simply a lifting of the degeneracy of
the single-triangle energy then the \sqtt\ pattern, with equal numbers of
$(+)$ and $(-)$ triangles, would have an energy $(F_{++}-F_{--})/2$.
Instead we find that $F_{-+}$ is a little lower than this, indicating that
the non-zero wire width also generates an interaction between the triangles
that favors the \sqtt\ pattern.  For wire widths near those of the
experiments, the single-triangle effect dominates over this
interaction, but for very narrow wires,
$w/a \cong 0.02$, the \sqtt\ pattern seems to have lower energy
than the \qzm\ pattern. 
We have not found an analytical understanding of this result for very thin wires.
It is somehow due to the nontrivial current pattern near the wire junctions.
 

For the experimental wire width of $w=0.1a$, we see from
Table \ref{table:London-1} that the non-zero wire width lowers the free
energy of the \qzm\ pattern by roughly $0.02E_L$ compared to a \sqtt\ or
random pattern.  For the experimental parameters we are using (Appendix A),
the London energy per unit cell is roughly
$E_L \cong 5000 k_B T_c (1-\frac{T}{T_c^m(f=1/2)})$. 
Thus if this were the only degeneracy-lifting effect, it would 
overcome the roughly $0.4k_B$ entropy difference per unit cell
and stabilize the conventional \qzm\ superconducting phase at
$(1-\frac{T}{T_c^m(f=1/2)}) \cong 0.004$.  This is below that phase's
KT temperature, so it would be also stable against vortices.  And this is
above the KT temperature of the $\psi^3$ ordered phase, so the system
will not enter this latter novel superconducting phase unless either this non-zero
wire width effect is reduced by making the wires narrower or changing the
experimental parameters or it is counterbalanced by some other effect.
What appears to be the simplest way of counterbalancing this effect is to
bend the wires slightly away from straight, while maintaining the full 
symmetry of the kagome grid, as is discussed below.


\subsection{Order-by-Disorder}

Another effect we consider that lifts the degeneracy in the minimum-energy
current patterns is the thermal fluctuations of $\psi$ around the minimum-energy 
patterns.
This effect is called 
``order-by-disorder'' \cite{villain}\cite{henley}\cite{chalker}. 
For the {\it kagome} antiferromagnet XY spin model (\ref{eq:antiferr}) with
nearest neighbor interaction and fixed-length spins, 
thermal fluctuations do not lift the degeneracy in the ground states \cite{huse} 
because they produce only phase fluctuations and these fluctuations are,
at lowest order, the same for all ground states.
But in our kagome-lattice superconducting wire network,
the magnitude of the order parameter can also fluctuate, and 
the free energy of these thermal
fluctuations does
lift the degeneracy among the current patterns in the Ginzburg-Landau
regime near $T_c$.

To estimate the effect of thermal fluctuations, we expand the energy 
of the system around the minimum-energy states, where 
$\frac{\delta F[\psi]}{\delta \psi^{\ast}}=0$, to quadratic order in 
the fluctuations.
The difference between the GL thin-wire limit free energy with and
without fluctuations is given by (\ref{eq:dev-energy}).
(We treat the fluctuations only in the thin-wire limit.)
Because of the coupling term between the magnitude and the phase
fluctuations, the energy of a given fluctuation
generally depends on the current pattern.
This is why the degeneracy in the minimum-energy current patterns 
can be lifted by thermal fluctuations.
Linearizing, the eigenmode $i$ of the fluctuations satisfies
\begin{eqnarray}
\frac{\delta F[\delta \psi_i]}{\delta (\delta\psi_i^{\ast})} 
&=&\Gamma_i \delta \psi_i,
\label{eq:TDGL}
\end{eqnarray} 
where $\Gamma_i$ is the stiffness of that eigenmode $i$.
We will focus on the soft modes, with small stiffnesses,
because these are the modes that have substantially different 
stiffnesses in the different patterns, thus lifting the degeneracy.
To lowest order in the amplitude of the fluctuations, their
contribution to the free energy is 
\begin{eqnarray}
\Delta F &=& \sum_{i} ~ \frac{k_B T}{2}  \log \Gamma_i 
\end{eqnarray}
(ignoring an additive constant which does not depend on the current
pattern). The sum runs over all eigenmodes of the fluctuations.
Like the other effects, we estimate 
this effect for the two extremal cases only: the \qz\ and \sqtt\ patterns.
Using the free energy difference (\ref{eq:dev-energy}) and equation
(\ref{eq:TDGL}), we obtain 
the eigenvalue equations for the magnitude $\delta|\psi|_i$ and the phase 
$\delta\phi_i$ of the fluctuations in eigenmode $i$:
\begin{eqnarray}
\Gamma_i \delta |\psi|_i & = & 4 |\psi_0|^2 \delta |\psi|_i 
+ 4 \xi_0^2 |\psi_0| \nabla_g \phi_0 \cdot \nabla \delta \phi_i 
- 2 \xi_0^2 \nabla^2 \delta|\psi|_i ~, \label{eq:TDGL-mag} \\
\Gamma_i |\psi_0| \delta \phi_i & = & -4 \xi_0^2 \nabla_g \phi_0 
\cdot \nabla \delta |\psi|_i -2 \xi_0^2 |\psi_0| \nabla^2 \delta \phi_i ~.
\label{eq:TDGL-pha}
\end{eqnarray}
These equations apply along each wire segment. The boundary conditions at 
each junction are that $\delta|\psi|_i$ and $\delta\phi_i$ are
continuous, and that the sums of the first derivatives of $\delta|\psi|_i$ and 
$\delta\phi_i$ moving away from each junction along each of the four wire segments 
meeting at that junction must both vanish.

\begin{figure}
\begin{center}
\leavevmode            
\epsfysize=5cm
\epsfbox{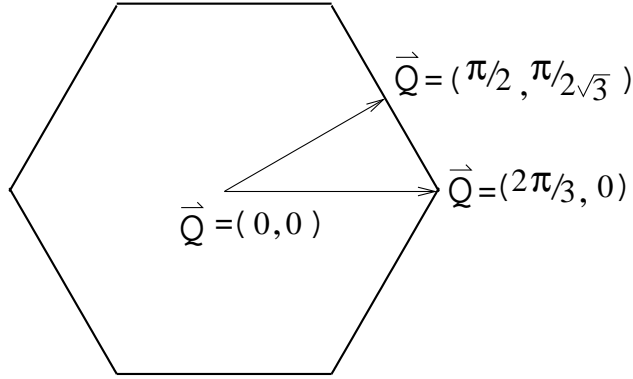}
\caption{First Brillouin zone of the kagome lattice.
}
\label{Brillouin_zone}
\end{center}
\end{figure}

For a thermal fluctuation with crystal momentum $\vec{Q}$, Bloch's theorem 
says
\begin{eqnarray}
\delta |\psi|(\vec{r}+\vec{R}) & = & e^{\imath \vec{Q} \cdot \vec{R}} 
\delta |\psi|(\vec{r})~, \label{eq:bloch-1} \\
\nabla \delta \phi(\vec{r}+\vec{R}) & = & e^{\imath \vec{Q} \cdot \vec{R}}
\nabla \delta \phi(\vec{r})~, \label{eq:bloch-2}
\end{eqnarray}
where $\vec{R}$ is any lattice translation vector connecting equivalent
points in two different unit cells. Using the equations (\ref{eq:bloch-1}) 
and (\ref{eq:bloch-2}), we can write the boundary conditions for the first 
derivatives of $\delta|\psi|$ and $\delta\phi$ at each junction in a unit 
cell as an eigenvalue equation with eigenvalue $\Gamma$:
\begin{eqnarray}
G_{\vec{Q}}(\Gamma)[\delta\psi]&=&0 ~, \label{eq:eigen}
\end{eqnarray}
where $G_{\vec{Q}}(\Gamma)$ is a $6\times 6$ matrix depending in
a nontrivial way on the eigenvalue 
$\Gamma(\vec{Q})$, and $[\delta\psi(\vec{Q})]$ is a column vector 
that consists of $\delta|\psi|$ and $\delta\phi$ at each of the three junctions 
in a unit cell. Our goal is to solve this eigenvalue equation for small 
$\Gamma(\vec{Q})$ at all points in the first Brillouin Zone (BZ)
(Fig. \ref{Brillouin_zone}). In general, this is not 
possible to do analytically, so we must solve for the
fluctuation spectrum numerically.  
However, near the mean-field transition temperature where $|\psi_0|$
is small we can obtain some analytic results that are perturbative
in $|\psi_0|$.

\subsubsection*{At the mean-field transition temperature}

At the mean-field transition temperature $T=T_c^m(f=1/2)$ there is only
one order parameter pattern that minimizes the GL free energy, namely the
normal state, with $\psi=0$ everywhere.  Here we can not solve for the
fluctuation eigenmodes in the gauge-invariant fashion discussed above,
because the phase $\phi$ is everywhere ill-defined.  However, it is
fairly straightforward to solve for the modes by choosing a convenient
gauge to work in.~\cite{thesis}
We find at $T_c^m(f=1/2)$ that the fluctuations have a complex zero-mode, with
stiffness $\Gamma_i=0$, over the entire first Brillouin zone.  Since
this zero mode is complex, it represents two real modes that will each
acquire a stiffness as soon as $|\psi_0|$ is
increased from zero by decreasing $T$ below $T_c^m(f=1/2)$.
The other modes all have stiffnesses of order one, except at
isolated points in the zone. 
In the GL limit of $|\psi_0| \ll 1$, it is only
the lowest two soft modes that contribute to the order-by-disorder
effect that lifts the degeneracy between current patterns, because
the higher modes have stiffnesses that do not depend on the patterns 
to lowest order in $|\psi_0|$.

\subsubsection*{In the limit  $|\psi_0|\rightarrow 0$}

When temperature decreases from the mean-field transition temperature,
the two degenerate zero modes that are present over the entire first
Brillouin zone for all current patterns acquire stiffnesses
of order $|\psi_0|^2$.  The modes at a given $\vec Q$ acquire stiffnesses
that depend on the current pattern.
Now we will examine these two 
lowest modes in order to estimate the free energy difference 
between the \qz\ and \sqtt\ patterns in this limit.

To obtain the stiffnesses of the soft modes with general momentum $\vec{Q}$ 
we expand the elements of our matrix $G_{\vec{Q}}(\Gamma)$
(\ref{eq:eigen}) 
to lowest order in $|\psi_0|$ for general $\vec{Q}$, using the fact that
the stiffnesses of interest are themselves of order $|\psi_0|^2$.
The stiffnesses of the two soft modes 
for the two patterns are, to lowest order in $|\psi_0|$,
\begin{eqnarray}
\Gamma_{\pm}[\sqrt{3}\times\sqrt{3}]&\approx& 2 |\psi_0|^2 [ \; 1 \pm 
\sqrt{1-\frac{6(3-w)}{33+16u+2v-2w}} \; ]~, \label{eq:Gamma-sqtt} \\
\Gamma_{\pm}[q=0]&\approx& 2 |\psi_0|^2 [\; 1 \pm
\sqrt{1-\frac{3(6-3u+w)}{21-10u+v+2w}} \; ]~; \label{eq:Gamma-q0} \\
u&\equiv&\cos({Q_x+\sqrt{3}Q_y})+\cos({2 Q_x})+
\cos({Q_x-\sqrt{3}Q_y})~, \nonumber \\
v&\equiv&\cos({2(Q_x+\sqrt{3}Q_y)})+\cos({4 Q_x})+
\cos({2(Q_x-\sqrt{3}Q_y)})~, \nonumber \\
w&\equiv&\cos({3Q_x+\sqrt{3}Q_y})+\cos({2\sqrt{3}Q_y})+
\cos({3Q_x-\sqrt{3}Q_y})~. \nonumber 
\end{eqnarray}
Because they are obtained ignoring the higher modes,
these analytic
expressions for the stiffnesses are not valid at all $\vec{Q}$:
For the \qz\ pattern the behavior is different at $\vec{Q}=0$,
where there are actually 4 soft modes. 
For the \sqtt\ pattern, there is a third soft mode at each of the
corners of the Brillouin zone.  However, in obtaining the order-by-disorder
effect to lowest order in $|\psi_0|$, these isolated special points in
the zone may be ignored, since they do not contribute to
the full free energy at that order.
The lowest-order fluctuation contribution to the free energy per unit cell
from the two soft modes is (again, ignoring an additive constant that is
independent of the current pattern)
\begin{eqnarray}
\Delta F &\approx& \frac{k_B T}{2} \frac{A_{hex}}{(2\pi)^2}
\int d^2 Q (\: \log \Gamma_+(\vec{Q}) + \log \Gamma_-(\vec{Q}) \:) .
\end{eqnarray}
In the limit of small $|\psi_0|$ the resulting contribution to the
free energy difference between our two current patterns is
\begin{eqnarray}
\Delta F_{\sqrt{3}\times\sqrt{3}}-\Delta F_{q=0}
&\cong& \frac{\sqrt{3}}{(2\pi)^2}(-3.68) k_B T \cong -0.16 k_B T~.
\end{eqnarray}
Thus in this limit there is a nonzero order-by-disorder effect, with the
\sqtt\ pattern having the lower free energy.

\subsubsection*{Intermediate region }

\begin{table}
\begin{center}
\caption{Stiffnesses of the four softest modes
for the \qz\ and \sqtt\ patterns 
at symmetry points in the first BZ at the KT temperature of the $\psi^3$
phase for our experimental parameters. All modes are obtained numerically.
The degeneracy of the modes are indicated in parentheses when they
are degenerate. }
\label{table:soft-finite-Q-2}
\vspace{.3cm}
\begin{tabular}{c|c|c} \hline \hline 
 $\vec{Q}$ & $\Gamma [q=0]$ 
& $\Gamma [\sqrt{3} \times \sqrt{3}]$  \\ \hline
(0,0) & 0, 1.02~~(2), 2.7~ & 0, 1.79~~(2), 2.7  \\ \hline
$(\frac{\pi}{2},\frac{\pi}{2\sqrt{3}})$ & 
0.35, 1, 1.4, 2.9 & 0.47, 0.7, 1.16, 2.8  \\  \hline
$(\frac{2\pi}{3},0)$ & 0.62~~(2),1.79 , 3. & 0.62~~(2), 1.02, $>$3. \\ 
\hline \hline
\end{tabular}
\end{center}
\end{table}

As the temperature is reduced, the order-by-disorder effect also
decreases, and it vanishes in the low-temperature London limit
of $\xi_0 \ll a$.
With our experimental parameters (Appendix A), the KT temperature of the 
$\psi^3$ phase falls in the crossover regime between Ginzburg-Landau
and London regimes.  Here we have to
solve the eigenvalue
equation (\ref{eq:eigen}) numerically for the stiffnesses $\Gamma(\vec{Q})$
of the fluctuation modes
with the experimental parameters. 
Table \ref{table:soft-finite-Q-2} shows the four softest modes at symmetry points
in the first BZ; we obtained these stiffnesses throughout the Brillouin zone.
We found that the free energy difference
between the two patterns is predominantly due to the
lowest three modes.
The resulting order-by-disorder contribution to the
free energy difference per unit cell between
the \sqtt\ and \qz\ patterns is roughly $0.04 k_B T$ 
at the KT temperature of the $\psi^3$ ordered phase, with the
\sqtt\ pattern still having the lower fluctuation free energy.
Note that this is indeed substantially lower than the roughly
$0.16 k_B T$ difference near the mean-field transition.  Note also that 
at this KT transition the
order-by-disorder effect is also much smaller than the finite-wire
width effect discussed above, so it is too weak to cause the
novel $\psi^3$ superconducting phase to be stabilized.


  

\subsection{Bending of the Wires}

From the previous two subsections, we know that
the effect of the non-zero wire width is dominant
over the order-by-disorder effect at the KT transition
of the $\psi^3$ ordered phase for the experimental
parameters, so the system orders into the \qzm\
pattern when the wire segments are straight. 
However, the novel $\psi^3$ 
superconducting phase might be stabilized
by bending the wires away from straight to
counteract the non-zero wire width effect.

\begin{figure}
\begin{center}
\leavevmode
\epsfysize=5cm
\epsfbox{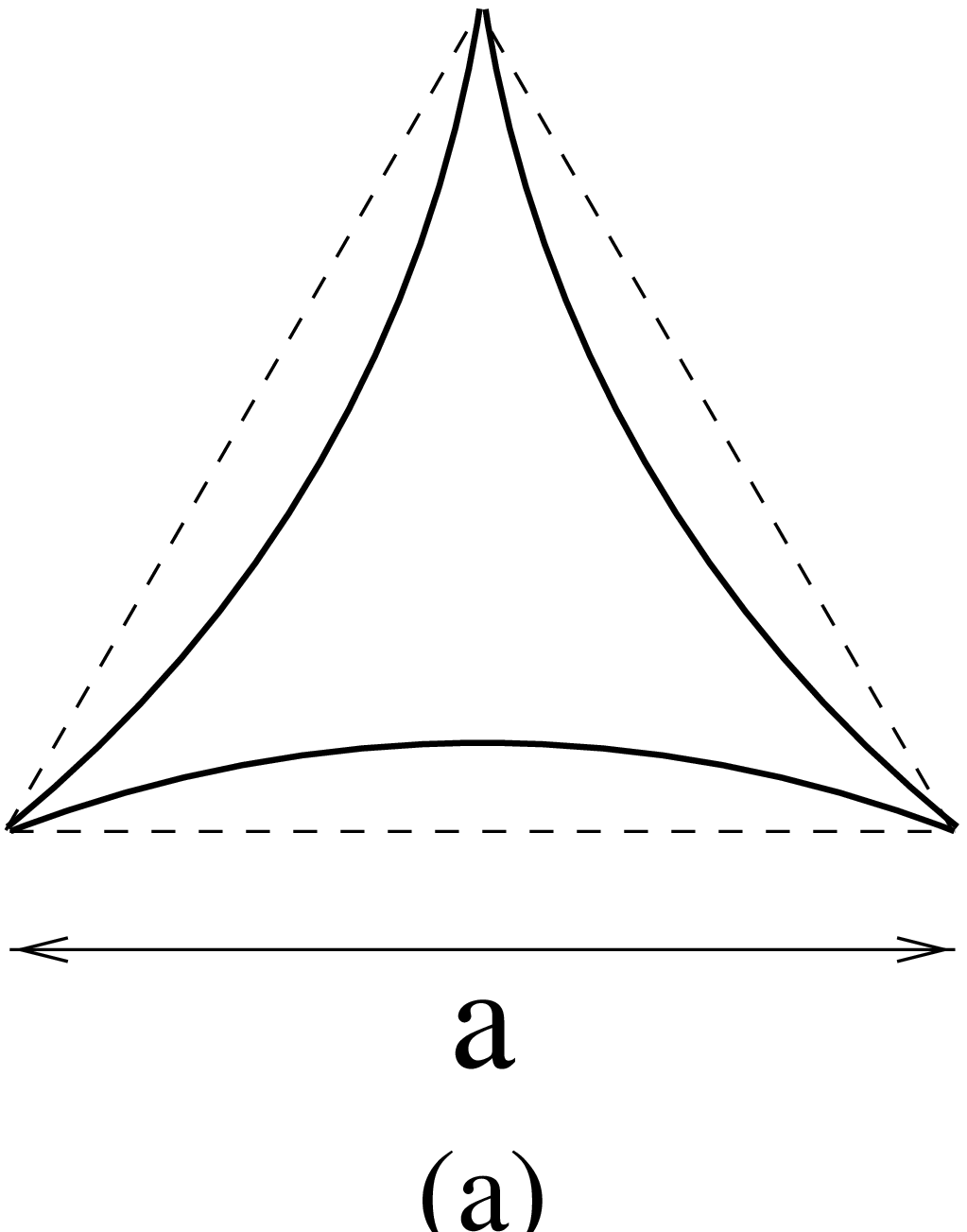}
\hspace{1.in}
\epsfysize=5cm
\epsfbox{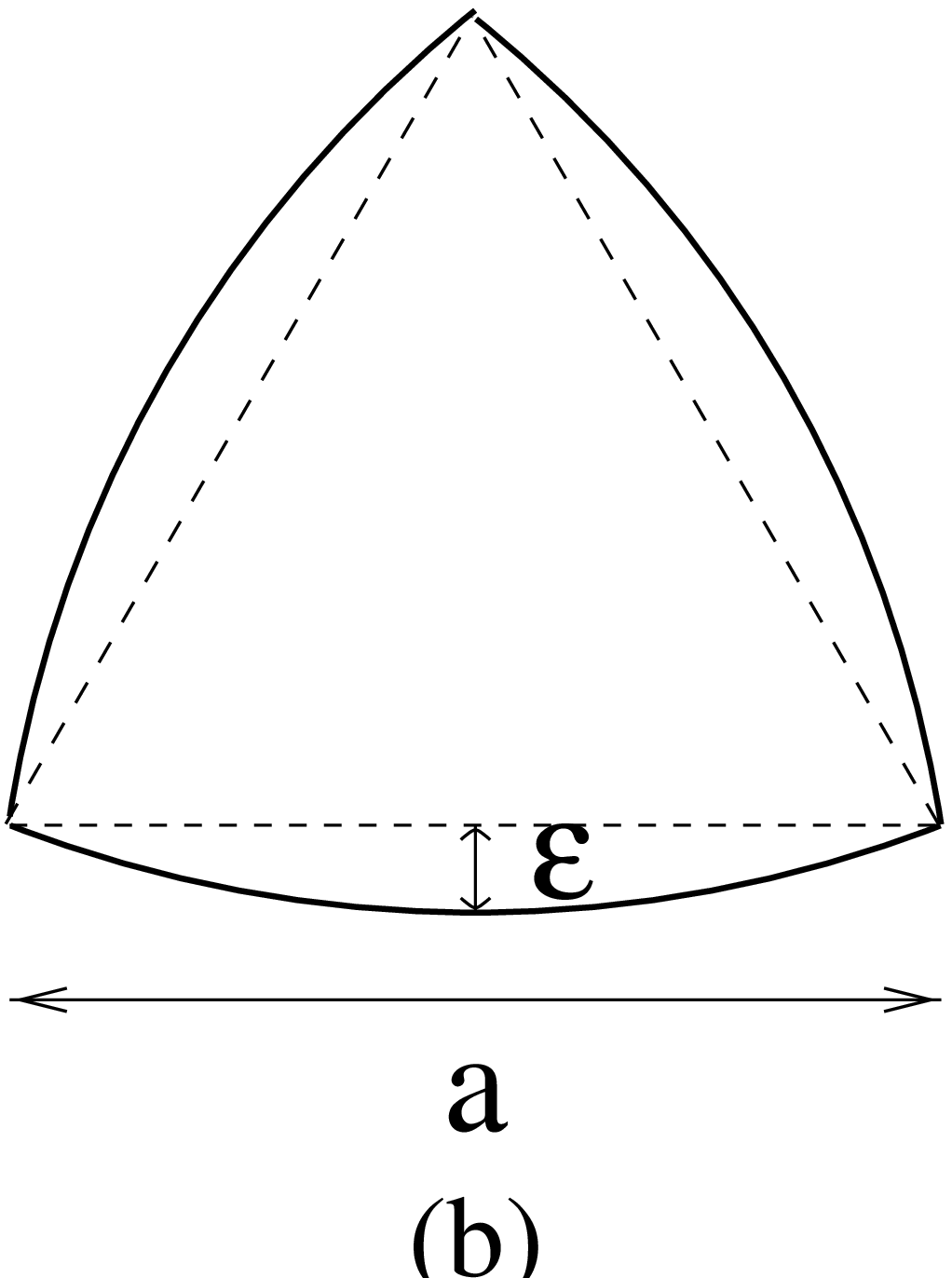}
\caption{Elementary triangles with the wires bent (a) in
towards the center of the triangle and (b) out away from
the triangle.
The straight lines between junctions are shown
dashed.}
\label{bending_wires}
\end{center}
\end{figure}

Let us consider a possible bending of the wires
away from straight while maintaining all translational
and rotational symmetries of the kagome grid.  We have been
calling the magnetic field of interest $f=1/2$, meaning one
half of a flux quantum through each triangle.  
We consider wire bending that
changes the area of a triangle, while not changing the area
of a full unit cell.  Thus once we bend the wires, the magnetic
field of interest is more precisely specified as being
four flux quanta per unit cell.  We consider bending the
wires by a distance $\epsilon$, while
keeping the junctions at their
original locations, as illustrated in Fig. \ref{bending_wires}.
The junctions are points of symmetry of the grid, which is
why they must stay put. 
This wire-bending lifts the degeneracy
of the two ground states of a single triangle.
The free energy of the $(-)$ pattern is lowered when the
wires are bent in towards the interior of the triangle,
and is increased when they are bent out as in 
Fig. \ref{bending_wires} (b).  Since the nonzero
wire width stabilizes the \qzm\ pattern, we are interested
in bending the wires out in order to destabilize it and allow
the system to order into the $\psi^3$ 
superconducting phase.

To analyze this quantitatively, we must assume some shape
for the bent wire segments.  For concreteness, let us assume they
are arcs of constant curvature.  Then the extra area of the triangle
due to the bending is simply $2a\epsilon$, to lowest order in
$\epsilon$.  (Negative $\epsilon$ corresponds to bending the
wires in.)




The resulting gauge-invariant phase gradients on the $(+)$ and $(-)$
triangles $|\nabla_g \phi_0^{(\pm)}|$ are, to first order
in $\epsilon/a$
,
\begin{eqnarray}
|\nabla_g \phi_0^{(\pm)}| &\approx & \frac{\pi}{3a}
(1 \mp \frac{8 \epsilon}{\sqrt{3} a}) ~. 
\end{eqnarray}
This results in a Ginzburg-Landau energy difference between the two patterns of
\begin{eqnarray}
E^{(+)} - E^{(-)} &\approx& -\frac{16 \epsilon}
{\sqrt{3} a} E_L~.
\end{eqnarray}
Bending the wires does not affect the energy of the
\sqtt\ pattern at order $\epsilon/a$.  For the experimental
parameters (Appendix A), to fully counteract the non-zero
wire width effect listed in Table \ref{table:London-1}
and bring the \qzm\ and \qzp\ patterns
back to degeneracy requires a surprisingly small bend of
$\epsilon \cong 70 \AA$.  With this $\epsilon$, the \sqtt\
pattern has the lowest free energy, 
mostly due to the non-zero wire width
effect, with the order-by-disorder effect also contributing.
For a smaller bend of roughly 20 to 50$\AA$, the \qzm\ pattern still has
the smallest energy, but by less than $0.4k_BT$; here
we expect the $\psi^3$ phase to be stabilized.  If the bend is increased
beyond about 90$\AA$, the \qzp\ pattern becomes the lowest energy.
Up to about 120$\AA$, the difference in energy between the \qzp\
and \sqtt\ patterns is too small to stabilize the \qzp\ pattern and
again we expect the $\psi^3$ phase to be stable.  Then the \qzp\
phase is stabilized for even larger bends.  Thus it appears that
any of the superconducting phases we have discussed can be stabilized
near the phase transition by making quite small bends
in the wires.  This strong sensitivity to wire-bending also indicates
that if random bends are present in the fabrication, this may introduce
rather strong quenched disorder that will locally favor specific
current patterns.

















\subsection{Summary}

\begin{figure}
\begin{center}
\leavevmode            
\epsfysize=8cm
\epsfbox{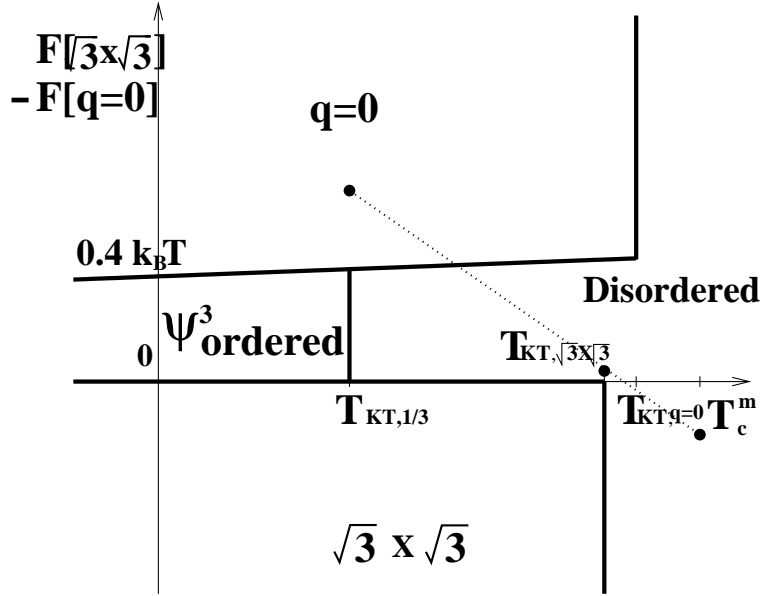}
\caption{ Rough phase diagram of the kagome 
superconducting wire network at $f=1/2$. 
The horizontal axis is the temperature.  
We show only the narrow temperature range close to 
the mean field transition temperature $T_c^m$, where the
Kosterlitz-Thouless transitions are.  
The vertical axis is the energy difference per unit cell 
between the \sqtt\ and the \qz\ patterns.
The boundaries of the three superconducting phases are shown bold. 
The dotted line shows the path taken in this phase diagram for the
experimental parameters that we consider (Appendix A).  This system
orders into a conventional superconducting phase with a $q=0$
current pattern.  As we discuss, the other phases can each be stabilized
by bending the wire segments by a suitably chosen amount.
}
\label{kg_phase_diagram}
\end{center}
\end{figure}

We summarize the degeneracy-lifting effects in the rough
phase diagram shown in Fig. \ref{kg_phase_diagram}.
For the parameters of the recent experiment \cite{yi}
with straight wires, the non-zero wire width effect is
dominant near the KT transitions
and causes the system to order into the \qzm\
phase.  However, as we discuss above, bending the wires a
little can counteract this effect and the amount of bending can be adjusted
to stabilize any of the
possible superconducting ordered phases.

This phase diagram (Fig. \ref{kg_phase_diagram}) is quite rough,
showing just the simplest estimates of the locations of the
phase boundaries.  Certainly the sharp corners shown in the
boundaries of the $q=0$ and \sqtt\ phases are unrealistic;
the true phase boundaries should be much smoother.  Also, one
should differentiate between the effects that lift the degeneracy
of single triangles and those that produce interactions
between triangles.  Thus there should be at least two axes of
degeneracy-lifting strengths.  We leave a more careful 
exploration of the statistical mechanics of this interesting
system for future researchers.

Another concern that we consider in the following
section is the kinetics of this system.
In order for this system to equilibrate, 
vortices must be able to move across the
wire segments, producing phase slip and changing the current pattern.
At too low temperature, these processes do not happen on laboratory
time scales, and the system is frozen into a particular
nonequilibrium current pattern.  Thus for our novel $\psi^3$
superconducting phase to be realized in the laboratory, its KT
transition temperature must be above the freezing temperature where
the system stops being able to equilibrate.














\section{Barriers to Vortex Motion}

In the novel $\psi^3$ superconducting phase, the kagome-grid
superconductor is not frozen into any one particular current pattern.
One thing that could cause a selection of one pattern is a
strong enough lifting of the degeneracy of the current patterns.
Let us now assume that the net degeneracy-lifting effect is
small enough that the equilibrium ordered phase is the 
$\psi^3$ phase.  However, to realize this phase experimentally
(or in a simulation), the system has to be able to equilibrate,
which in this case means to be able to fluctuate among the many
degenerate current patterns.  
The process by which the current pattern is
changed is that of phase-slip or vortex motion.  Here we
will describe the vortices of the $\psi^3$ phase and estimate
the barriers for their creation and motion in the vicinity
of this phase's KT transition temperature.  We find that the
barrier for creation of a vortex pair is roughly $16k_BT$, 
which certainly slows down the dynamics considerably, and 
makes proper equilibration quite a challenge for a simulation
study, but on the much larger time scales of experiment this
barrier will not stop equilibration.

\begin{figure}
\begin{center}
\leavevmode
\epsfysize=8cm
\epsfbox{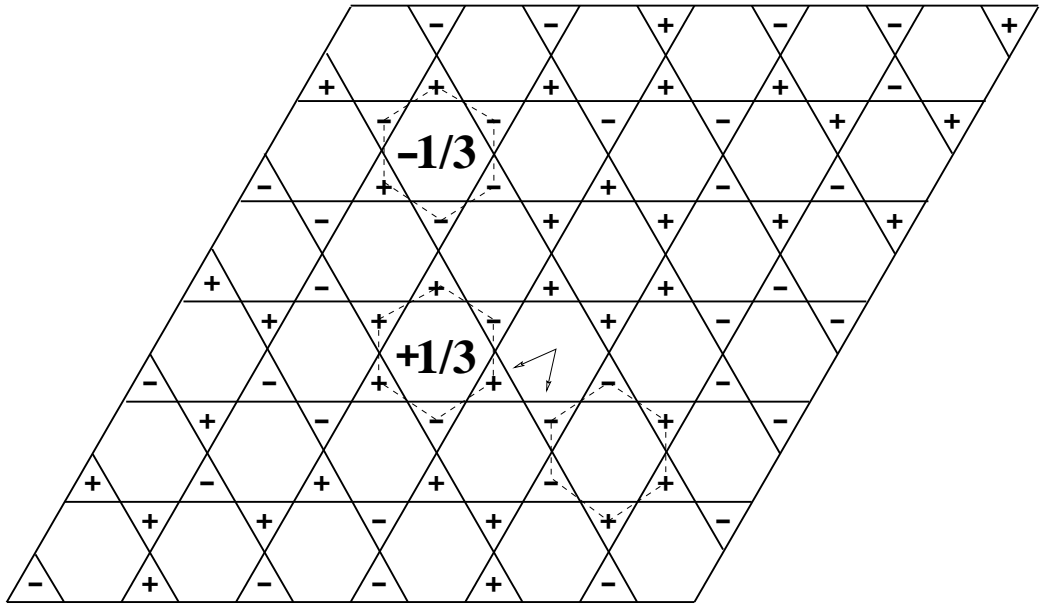}
\caption{A pair of $\pm 1/3$-vortices generated from a 
minimum-energy current pattern (Fig. \ref{kagome-patterns} (d)) 
through phase-slip processes
across the wires. The dotted lines represent hexagonal unit cells
where $\pm 1/3$ vortices are defined. As one example of
vortex motion, the $+1/3$ vortex can move 
to the empty hexagonal unit cell outlined by a dotted line
through phase slips across 
the wires indicated by the arrows. }
\label{frc-vortices}
\end{center}
\end{figure}

The vortices of $\psi^3$ that unbind at the KT transition of our
novel phase are 1/3-vortices of $\psi$.  The motion of these
vortices also causes rearrangement of the current pattern.
What do such fractional
vortices look like?  To make a pair of them, 
take a minimum-energy current pattern 
(for example, Fig. \ref{kagome-patterns} (d))
and move a conventional superconducting vortex from a $(+)$ triangle 
to an adjacent $(-)$ triangle 
through phase slips across the wires or at the intervening 
junction, as shown in Fig. \ref{frc-vortices}. 
This is equivalent to switching the circulations of these two adjacent
triangles. The gauge-invariant phase gradients $\nabla_g\phi$ around
the wire grids are rearranged such that the sum of $\nabla_g\phi$ 
around each triangle remains $\pm \pi$ and $\psi$ is single-valued.
This new current pattern is {\it not} one of the minimum-energy patterns.
This rearrangement 
creates a pair of $\pm 1/3$-vortices  
(see Fig. \ref{frc-vortices}) of the $\psi^3$ phase. 
To see these 1/3-vortices we use the hexagonal
unit cells shown dashed in Fig. \ref{frc-vortices}, whose corners are at 
the centers of the elementary triangles of the kagome grid.
The fractional vortices are different from ordinary vortices 
in the sense that the gauge-invariant vorticity $\oint (\nabla \phi - A)$ 
in a hexagonal unit cell may be an integer multiple of $2\pi/3$ instead of
the usual $2\pi$. 
Consider the $+1/3$-vortex in
Fig. \ref{frc-vortices}.  Since each triangle is shared by three hexagonal 
unit cells, we assume that each triangle contributes a third of 
its gauge-invariant vorticity of $\pm \pi$ to each of the three unit cells 
that overlap it.  For the unit cell containing our
$+1/3$-vortex, there are four adjacent
$(+)$ triangles and two $(-)$ triangles, and the gauge-invariant
vorticity in the elementary hexagon is zero.  Thus the net gauge-invariant 
vorticity in this hexagonal unit cell marked $+1/3$ is $+2\pi/3$.
With this definition of the vorticity in a hexagonal unit cell, all
minimum-energy current patterns are vortex-free, as should be the
case for the ground state of an ordered phase.
Once fractional vortices are present, they can also be moved by essentially
the same process that creates the pair.  For example, 
by switching the circulations of the two triangles which are indicated by
the arrows in Fig. \ref{frc-vortices} we can move the $+1/3$ vortex
to another location.  This illustrates how a
fractional vortex can move and how it rearranges the current pattern
as it moves. 

To estimate numerically the energy barriers to creating and moving
1/3-vortices, we need to find continuous paths in configuration space
for these processes. 
We do this using the GL thin-wire limit free energy (\ref{eq:1dim-GL-F})
on small systems with periodic boundary conditions, and discretized with
a finite number of numerical grid points along each wire segment.~\cite{thesis}
A continuous path from a given local minimum of the energy to a 
phase-slip event can be obtained by constraining the magnitude of the
order parameter at the location of the phase slip and relaxing all 
the other degrees of freedom to minimize the energy with this constraint.
Then the magnitude of the order parameter at the constrained point is
decreased to zero and we follow the energy during this process.  To find a 
path between two local minima that are simply separated by one phase-slip
event, we approach from each minimum, thus making the connection.
For the experimental parameters we use (Appendix A), 
and at the KT temperature of
the $\psi^3$ phase, we find that the lowest saddle point for the
fractional-vortex motion and pair creation processes is a phase-slip event
occurring at or very near a junction.  Thus it is the order parameter
magnitude at that junction that we suppress to find the saddle point.

Extrapolating our numerical
results to the continuum limit, we find that at the KT 
transition of the $\psi^3$ phase, the barrier for pair creation is
about $16k_BT$, and the barrier for motion of the $1/3$-vortices is
about $7k_BT$.   
This implies that the probability of creating a
1/3-vortex pair is roughly $e^{-16} \cong 10^{-7}$ per unit cell
per microscopic time scale.  Although this is a small number,
the size and time scales of the laboratory experiments 
are large enough that
this process will not be frozen out at $T_{KT,1/3}$, and the 
system can equilibrate into the novel $\psi^3$ phase. 
For simulation studies, however, this small number means equilibration
will be quite slow using standard simple algorithms \cite{binder}.  
This suggests that
simulation studies of this system should use special biased sampling
moves of some sort that will enhance the rate of these phase-slip
processes.

\section{Conclusion}

We have examined which superconducting state is stable among
the \qz,\ \sqtt,\ and novel $\psi^3$ ordered phases at
the Kosterlitz-Thouless (KT) transition in the kagome-lattice 
superconducting wire network at transverse magnetic field $f=1/2$. 
We estimated 
the helicity moduli of the \qz\ and \sqtt\ patterns
within the (straight) thin-wire GL theory without thermal 
fluctuations. In the GL regime, different minimum-energy 
current patterns have different helicity moduli 
because of the coupling between order parameter magnitude
and phase distortions. The helicity modulus of the \sqtt\ pattern 
is lower than that of the \qz\ pattern.
We use the helicity moduli to 
determine the KT transition temperatures for the $\psi$ ordered phases
and the novel $\psi^3$ phase.
At these KT temperatures we estimated the single-triangle 
degeneracy-lifting effects (non-zero wire width and
a possible bending of the wires) and the interactions 
between adjacent triangles (inductive coupling between 
adjacent supercurrents, order-by-disorder caused 
by thermal fluctuations, and also a non-zero wire-width contribution).
For the experimental parameters (Appendix A),
the system with straight wires becomes ordered into 
a conventional superconducting phase with the \qzm\ 
current pattern due to the dominant finite wire width effect.
But this finite wire width effect can be adjusted
experimentally by fabricating the wires narrower (with a fixed
lattice constant $a$) or, perhaps more easily, such that
they are bent away from the straight line between junctions
by an amount that restores the degeneracy.
Both ways can reduce the finite wire width effect so that
the system can be stabilized into the novel $\psi^3$ 
superconducting phase.
The energy barriers for creating and moving $1/3$-vortices
are low enough that the experimental system can equilibrate 
at and even a little below this KT transition.

\appendix
\begin{center}
{\Large \bf Appendix}
\end{center}

\section{Experimental Parameters}

For comparison to our analysis, we use the following
values of the parameters, as
in the recent experiment \cite{yi} on an aluminum kagome-lattice 
wire network:  
The number of unit cells is of order $10^5$. The distance between
junctions $(a)$ is $20000 \AA$. The width $(w)$ and thickness $(d)$
of the wires are $2000 \AA$ and $500 \AA$, respectively.
The zero-field transition temperature $T_c^m$ is $1.183K$.
The amplitude of the mean-field coherence length in zero field
scaled by $a$, $(\tilde{\xi}_0/a)$, can be estimated experimentally
by comparing the measured 
phase boundary $T_c^m(f)$ 
with the mean-field phase boundary obtained theoretically by Lin
and Nori \cite{yn}. Experimentally, it is found that the shape of
the $T_c^m(f)$ curve agrees well with the theory when a criterion
of some value between $1/20$ and $1/30$ of the normal state resistance 
is used to define $T_c^m(f)$.
This correspondence yields $(\tilde{\xi}_0/a) \cong 0.054$.
From the nonlinear current-voltage characteristics,
the KT temperature at $f=1/2$ is measured to be: 
$T_c^m(f=1/2)-T_{KT}(f=1/2) \cong 7.6 mK$.

Since the temperatures that we are interested in are all near
the zero-field transition temperature $T_c^m$, we use the following
approximate expressions \cite{tinkham} for the thermodynamic critical field
$H_c(t)$, the coherence length $\xi(t)$, and the penetration 
depth $\lambda(t)$, where $t$ is the reduced temperature $T/T_c^m$.
\begin{eqnarray*}
H_c(t)&\cong&1.73 H_c(0) (1-t)~, \\
\lambda(t)&\cong&\frac{\lambda_L(0)}{\sqrt{2(1-t)}} 
(\frac{\xi^{(cl)}(0)}{1.33 l})^{1/2}~, \\
\xi(t)&\approx&\tilde \xi_0(1-t)^{-1/2} \cong 
0.855 (\frac{\xi^{(cl)}(0) l}{1-t})^{1/2}~,
\end{eqnarray*}
where for aluminum samples \cite{roberts}\cite{kittel}, 
\begin{eqnarray*}
H_c(0) \cong 100 Oe,~~ \xi^{(cl)}(0) \cong 16000 \AA ,~~ \lambda_L(0) \cong 160 \AA,
~~ l \cong 100 \AA.
\end{eqnarray*}
Our estimate of the mean free path $l$ comes from the value $\tilde{\xi}_0
\cong 0.054 a$. The expressions for the coherence length and penetration 
depth are for the dirty limit, while $\xi^{(cl)}(0)$ is the low-temperature
coherence length in the clean limit.  The dirty limit applies to the samples
and temperatures we consider.   This sample falls in the type-II regime 
\begin{eqnarray}
\kappa|_{{\rm dirty~ limit}} &\cong& 0.715 \frac{\lambda_L(0)}{l},
\end{eqnarray}
although clean aluminum is a type-I superconductor.

\end{document}